\documentclass[a4paper,11pt]{article}
\pdfoutput=1
\usepackage{jcappub}
\usepackage[T1]{fontenc}
\usepackage{color}
\usepackage{amssymb}
\usepackage{amsmath}
\usepackage{upgreek}

\usepackage{xspace}
\usepackage{graphicx}
\usepackage{xcolor}
\usepackage{mhchem}
\usepackage{hyperref}
\usepackage{booktabs}
\usepackage{multirow}
\usepackage[mathlines,running]{lineno}

\newcommand{\nnbb}    {\ensuremath{2\nu\beta\beta}\xspace}
\newcommand{\onbb}    {\ensuremath{0\nu\beta\beta}\xspace}
\newcommand{\onbbM}   {\ensuremath{0\nu\beta\beta J}\xspace}
\newcommand{\onbbMM}  {\ensuremath{0\nu\beta\beta JJ}\xspace}
\newcommand{\onbbff}  {\ensuremath{0\nu\beta\beta\chi\chi}\xspace}
\newcommand{\nNbb}    {\ensuremath{\nu N\beta\beta}\xspace}
\newcommand{\Ge}      {\ce{^{76}Ge}\xspace}
\newcommand{\Xe}      {\ce{^{136}Xe}\xspace}
\newcommand{\kgyr}    {{kg\;yr}}
\newcommand{\qbb}     {\ensuremath{Q_{\beta\beta}}\xspace}
\newcommand{\gerda}   {\mbox{{\sc Gerda}}\xspace}

\newcommand{\mage}    {MaGe\xspace}
\newcommand{\PI}      {Phase~I\xspace}
\newcommand{\PII}     {Phase~II\xspace}
\newcommand{\alv}     {\ensuremath{\smash{\mathring{a}_\text{of}^{(3)}}}\xspace}
\newcommand{\sinT} {\ensuremath{\sin^2\theta}\xspace}
\newcommand{\pvalue}{\mbox{\textit{p}-value}\xspace}

\title{\boldmath Search for exotic physics in double-$\beta$ decays with \mbox{\protect{\sc{Gerda}}} \PII}

\author[1]{The \mbox{\protect{\sc{Gerda}}} collaboration,\note{email: gerda-eb{@}mpi-hd.mpg.de} } %
\author[j]{\\M.~Agostini,}
\author[j]{A.~Alexander,}
\author[v]{G.~Araujo,}
\author[p]{A.M.~Bakalyarov,}
\author[a]{M.~Balata,}
\author[n]{I.~Barabanov,}
\author[v]{L.~Baudis,}
\author[i]{C.~Bauer,}
\author[o,n,2]{S.~Belogurov,\note{\textit{also at:} NRNU MEPhI, Moscow, Russia}}
% ...
\affiliation[a]{INFN Laboratori Nazionali del Gran Sasso, Assergi, Italy}
% ...
\affiliation[i]{Max-Planck-Institut f{\"u}r Kernphysik, Heidelberg, Germany}
\affiliation[j]{Department of Physics and Astronomy, University College London, London, UK}
% ...
\affiliation[n]{Institute for Nuclear Research of the Russian Academy of Sciences, Moscow, Russia}
\affiliation[o]{Institute for Theoretical and Experimental Physics, NRC ``Kurchatov Institute'', Moscow, Russia}
\affiliation[p]{National Research Centre ``Kurchatov Institute'', Moscow, Russia}
% ...
\affiliation[v]{Physik-Institut, Universit{\"a}t Z{\"u}rich, Z{u}rich, Switzerland}

\abstract{A search for Beyond the Standard Model double-$\beta$ decay modes of \Ge has been performed with data collected during the \PII of the GERmanium Detector Array (\gerda) experiment, located at Laboratori Nazionali del Gran Sasso of INFN (Italy). Improved limits on the decays involving Majorons have been obtained, compared to previous experiments with \Ge, with half-life values on the order of 10$^{23}$\;yr. For the first time with \Ge, limits on Lorentz invariance violation effects in double-$\beta$ decay have been obtained. The isotropic coefficient \alv, which embeds Lorentz violation in double-$\beta$ decay, has been constrained at the order of $10^{-6}$\;GeV. We also set the first experimental limits on the search for light exotic fermions in double-$\beta$ decay, including sterile neutrinos.}

\begin{document}
\maketitle
\flushbottom

\section{Introduction}

The Standard Model (SM) of particle physics allows for double-$\beta$ decay, a nuclear transition in which the atomic number increases by two units with the number of nucleons staying constant, as long as two anti-neutrinos and two electrons are also emitted, $(A,Z) \rightarrow (A,Z+2) + 2e^- + 2 \Bar{\nu}$. This is called two-neutrino double-$\beta$ (\nnbb) decay and is one of the rarest radioactive processes ever observed~\cite{Barabash:2020nck}. Double-$\beta$ decays are powerful probes of physics Beyond the Standard Model (BSM). The most popular BSM double-$\beta$ decay mode in which no anti-neutrinos are emitted, the so-called neutrinoless double-$\beta$ (\onbb) decay, violates lepton number and, if observed, would imply that neutrinos have a Majorana mass component~\cite{Schechter:1981bd}. The existence of new particles in BSM theories can lead to double-$\beta$ decay modes with different final states, in which the exotic particles are emitted along with the two electrons. More exotic double-$\beta$ decay modes can also be considered, in which the final state is the same as in \nnbb decay, but the decay kinematics is affected by the BSM physics effect. 

\begin{figure}[tbp]
\centering 
\includegraphics[width=.75\textwidth]{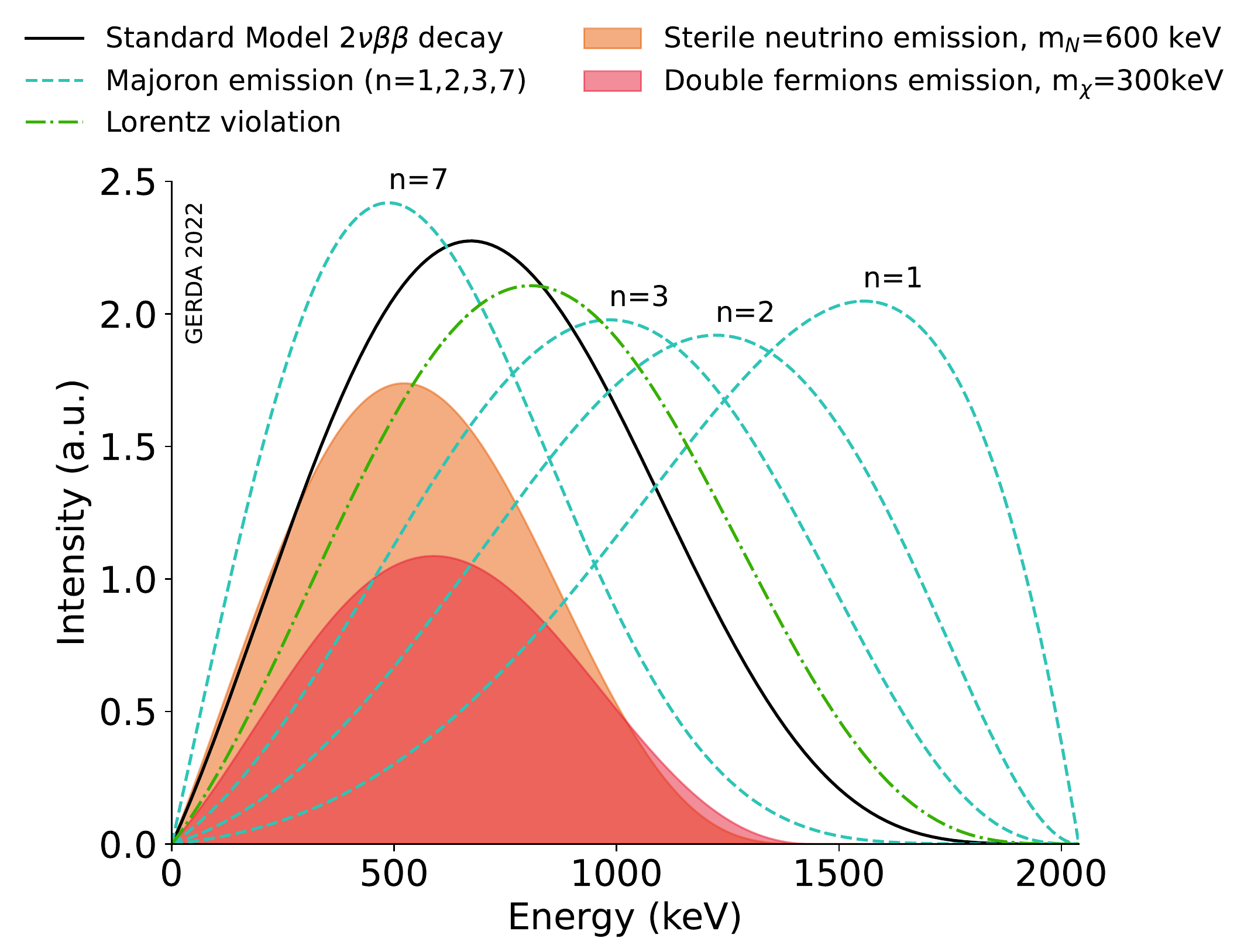}
\caption{\label{fig:shape_doublebeta}Energy distribution of different BSM double-$\beta$ decay modes shown in comparison to the SM \nnbb decay distribution. The normalization of the distributions is arbitrary and adjusted for better visualization. }
\end{figure}

The experimental quantity that allows distinguishing exotic decays from the SM \nnbb decay is the shape of the predicted distribution of the summed energy of the two emitted electrons. The shape for \onbb decay, in which the two electrons carry all the available decay energy, the Q-value (\qbb) of the decay, is simply a peak at this energy value, broadened by the experimental energy resolution. Decays that involve other particles in the final state, which escape detection by the experiment, like the anti-neutrinos in the SM \nnbb decay or other exotic particles, are characterized by a continuous distribution between 0 and \qbb. As a first approximation, the shape of this distribution can be characterized by a spectral index $n$, that appears in the phase space of the decays $G\sim(Q_{\beta\beta}-E)^n$, where $E$ is the summed energy of the two electrons. For instance, the SM \nnbb decay shape is defined by spectral index $n=5$. Massive exotic particles can also be produced in double-$\beta$ decay if kinematically allowed, i.e.~if their mass is smaller than the \qbb. Also in this case, a continuous spectrum is expected, whose end-point is now shifted at $(Q_{\beta\beta}-m_X)$, where $m_X$ is the mass of the exotic particle $X$. In figure~\ref{fig:shape_doublebeta} the energy distribution predicted for different BSM double-$\beta$ decay modes is shown compared to the SM \nnbb decay distribution.  

In this paper, we report on the search for new physics with \Ge double-$\beta$ decays performed in \gerda \PII, in the energy region dominated by \nnbb decay, between 560\,keV and 2000\,keV. Results on the search for \onbb decay with the full \gerda data set have been published in~\cite{GERDA:2020xhi}. Here we focus on BSM processes that manifest as a deformation of the continuous \nnbb decay spectrum, such as decays involving Majorons, light exotic fermions, and Lorentz violation.

The \gerda experiment is briefly described in section~\ref{sec:gerda_exp}. 
In section~\ref{sec:data_sel}, we discuss the data selection criteria and the methods used to model signal and background events in the analysis. The statistical analysis and the systematic uncertainties that contribute to this analysis are discussed in section~\ref{sec:stat&sys}. 
In section~\ref{sec:results}, we present our results on the search for exotic double-$\beta$ decay modes. The search for neutrinoless double-$\beta$ decay with Majorons emission is discussed in~\ref{sec:results_majorons}, the search for Lorentz and CPT violation in~\ref{sec:results_lorentz}, and the search for light exotic fermions in~\ref{sec:results_massivefermions}. Finally, we summarize our conclusions in section~\ref{sec:conclusions}.   

\section{The \gerda experiment}\label{sec:gerda_exp}
The \gerda experiment was located at the Laboratori Nazionali del Gran Sasso (LNGS) of INFN, in Italy, where a rock overburden of 3500\;m water equivalent reduces the flux of cosmic muons by six orders of magnitude~\cite{GERDA:2012qwd}. High-purity germanium (HPGe) detectors, isotopically enriched in \Ge, were operated inside a 64\;m$^3$ liquid argon (LAr) cryostat~\cite{Knopfle:2022fso}. 

In the second phase of the experiment, 10 coaxial (including 3 detectors with natural isotopic abundance) and 30 Broad Energy Germanium (BEGe) detectors were used~\cite{GERDA:2017ihb}. After an upgrade in May 2018, the three natural coaxial detectors were removed, and 5 additional inverted coaxial (IC) detectors were installed. Detectors were mounted on 7 strings, and each string was placed inside a nylon cylinder to limit the collection of radioactive potassium ions on the detector surfaces~\cite{Lubashevskiy:2017lmf}. 

The LAr volume around the detectors was instrumented with a curtain of wavelength-shifting fibers connected to silicon photo-multipliers (SiPM) and 16 cryogenic photo-multiplier tubes (PMTs) to detect scintillation light in the LAr~\cite{Janicsko-Csathy:2010uif, GERDA:2017ihb}. During the upgrade, the geometrical coverage of the fibers was improved, more SiPM channels were added, and their radiopurity increased.

The cryostat was surrounded by a water tank containing 590\;m$^3$ of pure water, equipped with PMTs to detect the Cherenkov light of residual cosmic muons reaching the detector site. The instrumented water tank formed, together with scintillator panels on the top of the experiment, the muon veto system~\cite{Freund:2016fhz}.  

Data have been processed following the procedures and digital signal processing algorithms described in~\cite{Agostini:2011mh}. The energy of an event is reconstructed using a zero-cusp-area filter~\cite{GERDA:2015rik}. Weekly calibration runs with $^{228}$Th sources are performed to determine the energy scale and resolution. A resolution better than 3\;keV(4\;keV) full-width half maximum is achieved with BEGe and IC(coaxial) detectors over a wide energy range extending up to \qbb~\cite{GERDA:2021pcs}. 

\section{Data selection and modeling}\label{sec:data_sel}
The data used for this analysis were collected between December 2015 and April 2018, before the upgrade of the experimental setup in the summer of 2018. Of the 40 HPGe detectors deployed in the \gerda array during this period, only data collected with the 30 BEGe detectors have been used for this analysis. The corresponding data set exposure is 32.8\;\kgyr.
This data set was selected among the full exposure to minimize the systematic uncertainties. In fact, the BEGe detectors are better understood and have been characterized more accurately than the coaxial detectors, and the LAr veto instrumentation underwent major changes during the upgrade that are not included in the modeling of the LAr veto system~\cite{Agostini2022b}. The information of the LAr veto system and its Monte Carlo simulation are crucial elements of the analysis, as will be explained in the following.

The event topology of all the considered double-$\beta$ decays is the same, and it is a localized energy deposition within one germanium detector. The total decay energy is shared among the two electrons and either the two anti-neutrinos or one or more exotic particles produced in the process. The electrons release all their energy within a few millimeters from the decay vertex in germanium. Both anti-neutrinos and the exotic particles escape the detector carrying away part of the decay energy. Thus, in all the considered decays, the reconstructed energy varies between zero and the Q-value of the decay, $\qbb = 2039.061(7)$\;keV~\cite{PhysRevC.81.032501}.\footnote{If a massive exotic particle is produced in the process, the maximum energy is shifted to a lower value by the mass of the particle for a single production or twice the mass for pair production.}
Several background sources are also expected to generate events in this energy region~\cite{GERDA:2019cav}. While an energy deposition of a double-$\beta$ decay is fully contained in one germanium detector, $\gamma$ radiation, mostly interacting via Compton scattering, can undergo multiple separated energy depositions in more than one germanium detector. To reduce the $\gamma$ background, an anti-coincidence cut between different germanium detectors is applied. In addition, $\gamma$ rays can release part of their energy in the LAr volume surrounding the detector array and trigger the LAr veto system. An anti-coincidence cut between germanium detectors and the LAr veto system is also applied (LAr veto cut). This reduces the $\gamma$ background by a factor of $\sim$10 in the energy region dominated by the \nnbb decay,
as it was shown in~\cite{Agostini2022b},
while double-$\beta$ decay events are accepted with an efficiency of about 98\%, estimated from the accidental coincidences of randomly triggered events. Events preceded by a trigger in the muon-veto system within 10\,$\mu$s are also discarded, with negligible induced dead time (<0.01\%). In addition, each event has to pass several quality cuts based on the flatness of the baseline, polarity, and time structure of the pulse to reject non-physical events. The acceptance efficiency of physical events by quality cuts is larger than 99.9\%. The pulse shape discrimination (PSD), successfully used in the search for \onbb decay~\cite{GERDA:2020xhi}, was not used in this analysis. In fact, despite the possibility of discriminating between single-site signal events and multi-site background events, the efficiency of the PSD cut below 1\,MeV is poorly understood~\cite{Agostini2022}, and its energy dependence could introduce distortion of the energy spectrum thus, additional systematic uncertainties in the analysis.

The energy range chosen for the analysis extends from 560\;keV to 2\;MeV. At low energy, the event rate of \gerda is dominated by the decay of $^{39}$Ar, a cosmogenically produced argon isotope. Above 2\;MeV, the contribution due to \nnbb decay or any other exotic decay vanishes. 

\begin{figure}[tbp]
    \centering
    \includegraphics[width=\textwidth]{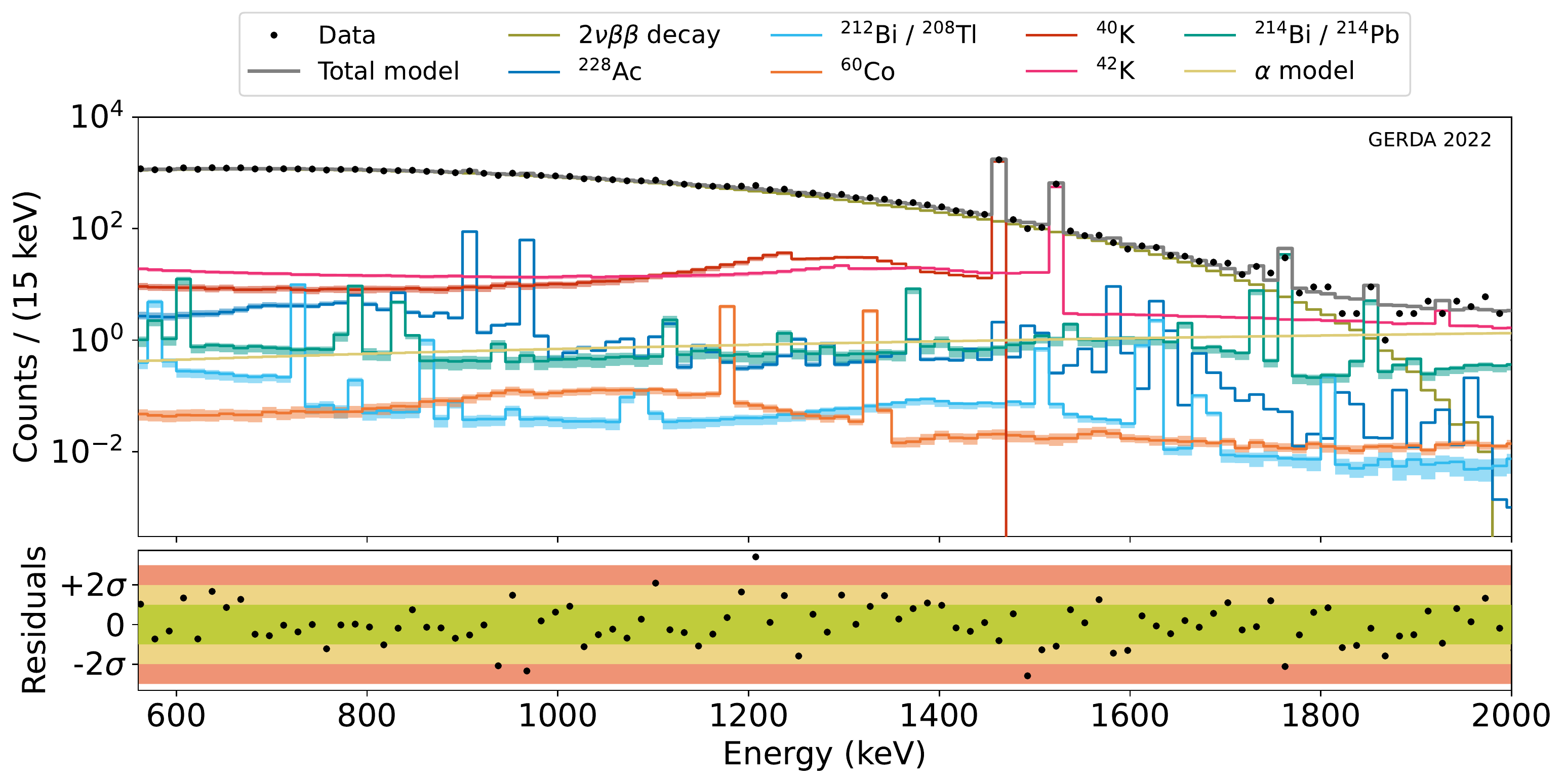}
    \caption{Background decomposition of the 32.8\,\kgyr\ of BEGe data from \gerda \PII after the application of the LAr veto cut, as derived in~\cite{Agostini2022b}. The shaded bands represent the uncertainty of the LAr veto modeling. In the bottom panel, the difference between data and the total model normalized over the expected statistical fluctuation in each bin is shown.}
    \label{fig:data_larmodeling_begeonly}
\end{figure}

The background model used in this analysis is a simplified version of the background model derived in \cite{GERDA:2019cav}. In the latter, where data before the LAr veto cut are used, contributions for a given background source at different locations are treated independently. In this analysis, where the LAr veto cut is applied, only a subset of these components is used. The expected background model after LAr veto cut, as derived from the modeling of the LAr veto in~\cite{Agostini2022b}, is shown in figure~\ref{fig:data_larmodeling_begeonly} for the 32.8\,\kgyr\ of BEGe data from \gerda \PII.
Above the Q-value of $^{39}$Ar $\beta$ decay at 565 keV, the dominant contribution to the energy spectrum is due to \Ge \nnbb decay. Minor contributions are due to $^{228}$Ac, $^{228}$Th, $^{214}$Bi, $^{60}$Co and $^{40}$K decays in structural materials. One component for each source is included in the background model, with the source of the decay located close to the detector array. $^{42}$K decays in LAr are modeled with two independent components, one relative to decays in the LAr volume and one to decays very close to the detectors' surfaces. Finally, a minor contribution to the background budget is expected from $\alpha$ decays on the p$^+$ electrode of the detectors. These decays do not generate detectable scintillation light, and no significant differences are therefore expected from the model in~\cite{GERDA:2019cav}. The distribution of $\alpha$ decays in the energy range of this analysis is well approximated by a linear function.

The probability distributions of signal and background events are obtained from Monte Carlo simulations, performed using the \mage simulation framework~\cite{Boswell:2010mr}, as detailed in~\cite{GERDA:2019cav}. Finite energy resolution, detector active volume model, and individual exposure of the detectors are considered. The simulation of the LAr scintillation light production and detection chain has also been implemented in the same framework~\cite{Agostini2022b}. This is used to determine the effect of the LAr veto cut on the probability distribution functions, not included in previous works~\cite{GERDA:2019cav}.

\section{Statistical analysis and systematic uncertainties}\label{sec:stat&sys}

To search for exotic decays, a binned maximum-likelihood fit is performed in the energy window between 560\;keV and 2000\;keV, using a 10\;keV binning. It was checked that the fit results are stable for different sizes of the binning. Assuming that the number of events in each bin is Poisson distributed, the likelihood function is given by the product of Poisson probabilities for all bins. The likelihood is then used to construct a frequentist test statistic based on the profile likelihood ratio~\cite{Cowan:2010js}. 

The distribution of the test statistic is evaluated with Monte Carlo techniques for the background-only hypothesis, corresponding to the parameter of interest being equal to zero, and for a discrete set of non-zero values of the parameter of interest. The test statistic distributions are used to extract the \pvalue of the data and the experimental sensitivity~\cite{Agostini:2017iyd}. 

The parameter of interest on which the statistical inference is made depends on the considered new physics process. In the search for \onbb decay with the emission of Majorons or pairs of light exotic fermions, the parameter of interest is the strength of the exotic decay signal, proportional to the inverse of the decay half-life and expressed in number of events in the fit range. In the search for sterile neutrinos and Lorentz-violating effects, the parameter of interest is an internal parameter that defines the \nnbb decay distribution and is responsible for its distortion compared to the SM prediction. Additional parameters of the fits are the number of background and \nnbb decay events observed in the fit range. These are treated as nuisance parameters and left unconstrained, and their uncertainties are propagated by profiling~\cite{Zyla:2020zbs}. 

Systematic uncertainties that affect the signal and background energy distribution and, in turn, the fit result are folded into the analysis during the computation of the test statistic distribution. This procedure implies a broadening of the test statistic distribution that is used for statistical inference. Consequently, systematic uncertainties are naturally incorporated in the result~\cite{Zyla:2020zbs}. 

Uncertainties on the signal and background shape distributions are generally related to the modeling of the background and the detector response. Given a $\gamma$ decay in the vicinity of a detector, this can deposit the full energy in the detector, resulting in a peak in the energy spectrum, or undergo Compton scattering, resulting in a continuous energy distribution up to the Compton edge. The ratio between the two classes of events is expected to change for $\gamma$ decays very close to the detector or far from it, as well as for different models of the LAr veto response. 

Different locations of the background sources have been identified in~\cite{GERDA:2019cav}, but this picture is changed by the LAr veto cut since the suppression efficiency is expected to be different for the different background components. On the other hand, without the introduction of prior knowledge, the low background level does not allow to distinguish between different locations of a background source in the fit. As previously mentioned, the fit model contains only one component for each isotope, except for $^{42}$K, for which two components are used to separate the contributions from the decays near and far from the surface of the detectors.
The systematic uncertainty introduced by this choice is accounted for in the Monte Carlo generation of the pseudo-experiments used to compute the test statistic distribution. When the pseudo-experiments are generated, the location of each background contribution is uniformly sampled among the full set of locations identified in~\cite{GERDA:2019cav}.

The modeling of the LAr veto response is affected by uncertainties in the optical parameters used in the Monte Carlo simulations, such as the LAr attenuation length and the reflectivity of different materials in the detector array, among others. A complete treatment of this topic can be found in~\cite{Agostini2022b}, where the methodology used to model systematic uncertainties of the LAr veto response is discussed. This model is used in the Monte Carlo generation of the pseudo-experiments to account for the systematic uncertainty affecting the LAr veto response.   

Modeling of germanium detector response can lead to additional systematic uncertainty. In the fully-active volume of the detector, where the charge collection efficiency is maximal, the deposited energy is always well reconstructed. The charge collection efficiency degrades in a $\sim$1\;mm deep transition region at the n$^+$ contact lithiated surface~\cite{Lehnert2016}. Energy deposition in this region is only partially reconstructed, depending on the efficiency profile and the size of the transition region. 
In turn, both these parameters affect the energy distribution of different fit components, in particular the lower tail of intense $\gamma$ peaks and the low energy region of the \nnbb decay. 
The charge collection efficiency is assumed to decrease linearly in the transition region, whose average size for the BEGe detector is about 50\% of the full dead layer region~\cite{Lehnert2016}. In the probability distributions, the transition layer size is assigned individually to each detector. In the Monte Carlo generation of the pseudo-experiments, the transition layer size of each detector is varied in a conservative range of $\pm$5 standard deviations from the central value.

In all the previous discussions, we referred to the SM \nnbb decay as the transition to the ground state of the daughter nucleus. In addition to the ground state, the \nnbb decay of \Ge can also proceed to the excited states of $^{76}$Se. To date, none of these excited state transitions have been observed with \Ge. Limits on the decay rate to the first three excited states have been set in \gerda \PI~\cite{GERDA2015}, and more recently by the \textsc{Majorana Demonstrator}~\cite{MAJORANA2020}. Among these excited states transitions, the decay to the $0^+_1$ state is expected to dominate, with the most recent half-life prediction of $(2.6\pm0.13)\cdot 10^{24}$\,yr~\cite{Kostensalo2022}. The experimental lower limit on the half-life of this process lies at $7.5\cdot 10^{23}$\,yr~\cite{MAJORANA2020}. If the \nnbb decay of \Ge to the $0^+_1$ excited state happens with a half-life of the order of $10^{24}$\,yr, it could introduce a systematic uncertainty in the search for additional sub-dominant contributions to the energy spectrum. Nevertheless, this decay is accompanied by the emission of two $\gamma$s with energies of 563.2\,keV and 559.1\,keV, which can be detected with high probability in the surrounding detectors or in the nearby LAr volume producing an event topology that does not survive the data selection discussed above. For this reason, the \nnbb decay of \Ge to the $0^+_1$ excited state was not considered in the background model. The systematic uncertainty in the search for exotic double-$\beta$ decays was estimated to be negligible. A more quantitative justification is presented in Appendix~\ref{app:excited_states}. 

Finally, uncertainties in the theoretical calculations of the shape of \nnbb decay are considered. Different shapes of \nnbb decay are predicted if the calculations assume the Higher-State Dominance (HSD) or the Single-State Dominance (SSD) model.
The HSD model is based on the assumption that all the intermediate states of the intermediate nucleus contribute to decay rate~\cite{Kotila:2012zza}. This hypothesis has always been assumed in the calculations of the shape of the \Ge \nnbb decay. The SSD model, on the other hand, assumes that the \nnbb decay is governed by a virtual two-steps transition through the first 1$^+$ state of the intermediate nucleus~\cite{Domin2005}. This model has been observed to describe better the \nnbb decay of several nuclei~\cite{Azzolini2019b, NEMO-3:2019gwo, Armengaud2020}. In \Ge, the difference between the two models is maximal (about 10\%) in the tail of the distribution, but less than 0.5\% around the peak of the distribution.\footnote{Private communications with J.~Kotila and F.~Iachello.} 
In both cases, the calculations have been performed using exact Dirac wave functions with finite nuclear size and electron screening as described in~\cite{Kotila:2012zza}. The \nnbb distribution is equally sampled from both models in the Monte Carlo generation of the pseudo-experiments.

\section{Results}\label{sec:results}
\subsection{Search for Majoron-involving decays}\label{sec:results_majorons}

Many theories of BSM physics predict the emission of massless or light bosons, the so-called Majoron~($J$), in double-$\beta$ decay. In early models, the Majoron is introduced as a Goldstone boson arising from the spontaneous breaking of the global $B-L$ symmetry~\cite{PhysRevLett.45.1926, Gelmini:1980re, Georgi:1981pg, Mohapatra:1988fk}. Precision measurements of the width of the $Z^0$ boson decay to invisible channels~\cite{ALEPH:2005ab} ruled out these old models. In the following years, new models were proposed in which the term Majoron is used in a more general sense to refer to a massless boson that couples with neutrinos and gives rise to neutrinoless double-$\beta$ decay accompanied by Majoron emission (\onbbM decay), $(A,Z) \rightarrow (A,Z+2) + 2e^- + J$~\cite{Berezhiani:1992cd, Burgess:1992dt, Carone:1993jv}. The emission of two Majorons is also predicted in some models (\onbbMM decay), $(A,Z) \rightarrow (A,Z+2) + 2e^- + 2 J$~\cite{Bamert:1994hb}. Following the classification in~\cite{Hirsch_1996, Kotila_2021}, the massless Majorons models can be grouped into 4 different classes according to the spectral index $n=1,2,3\text{ or }7$, which defines the energy distribution of the decay. The energy distributions for different \onbb decays with Majorons emission in \Ge are shown in figure~\ref{fig:shape_doublebeta}, compared to the distribution of the SM \nnbb decay.   

To search for \onbb decay with the emission of Majorons, the energy distribution of the decay is added to the fit as an independent component that constitutes the signal.
The parameter of interest in this analysis is the number of signal events observed in the data set, which is proportional to the inverse of the decay half-life.

\begin{figure}
    \centering
    \includegraphics[width=\textwidth]{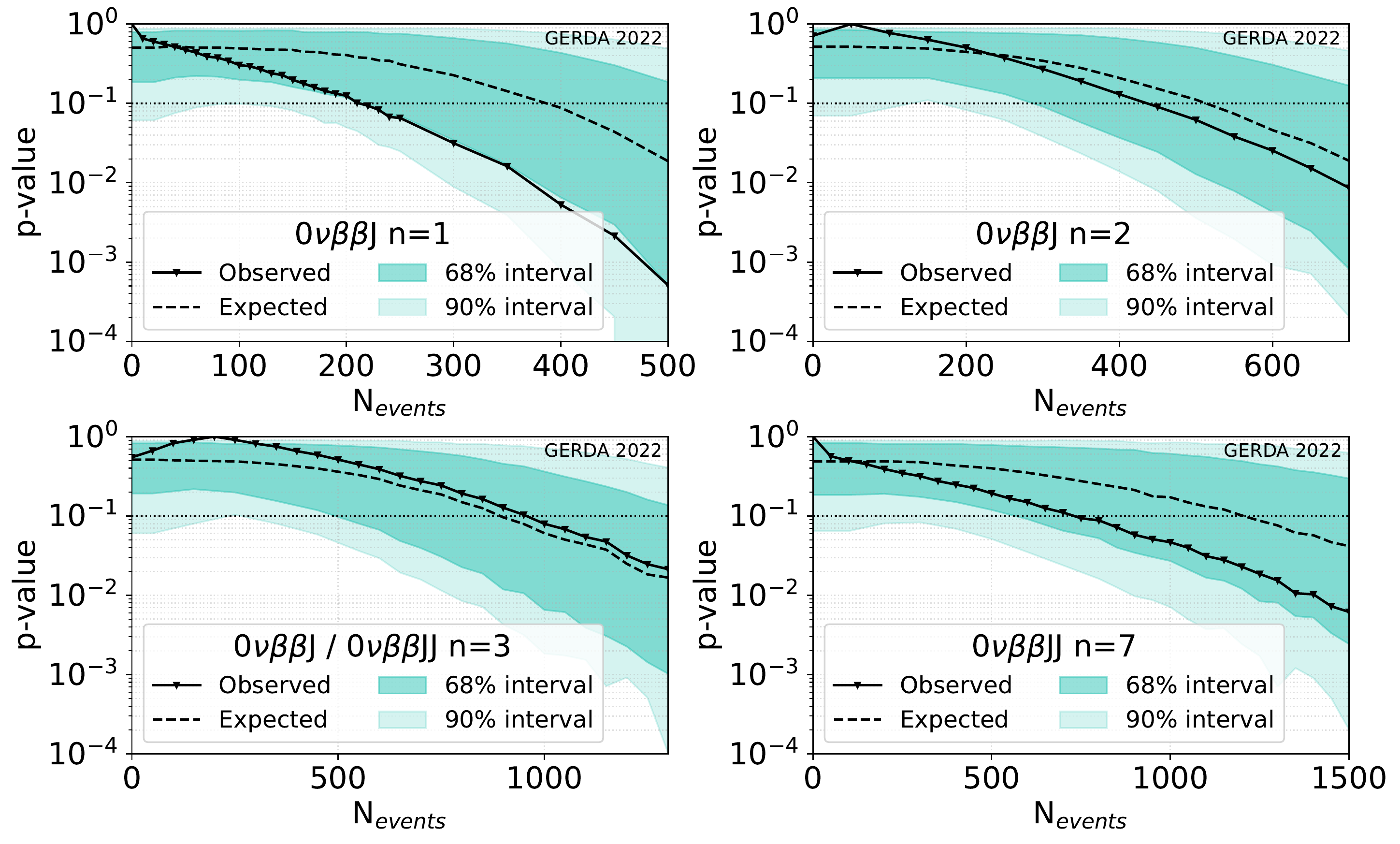}
    \caption{\pvalue as a function of the number of events in the fit range for the indicated Majoron decay modes. The dashed black line shows the median of the \pvalue distributions for several Monte Carlo realizations of the \gerda experiment with no signal. The colored bands indicate the spread of the distributions, given by the central 68\% and 90\% probability intervals. The observed \pvalue for the \gerda data is represented by the solid black line. The 90\% C.L.~limit (sensitivity) on the number of events in the fit range is given by the intersection of the solid (dashed) black line with the dotted line, corresponding to a \pvalue of 0.1.}
    \label{fig:majo_only}
\end{figure}

No evidence of a positive signal is found for any of the considered Majorons models. In the analysis data set, 46430 events are associated with the SM \nnbb decay, whereas 4610 events are associated with the other backgrounds. The best-fit value for the $n=1$ and $n=7$ models lies at zero. A non-zero best-fit value is found in the case of $n=2$ and 3, but the 68\% C.L.~interval includes zero. Therefore, a 90\% C.L.~limits has been set on the strength of each considered decay. The observed \pvalue has been evaluated for a discrete set of values of the parameter of interest, using the test statistic distribution computed with Monte Carlo methods. The resulting \pvalue distributions are shown in figure~\ref{fig:majo_only} for all the considered Majoron-involving decays. The median of the distribution of the expected \pvalue, which gives the experimental sensitivity, and the 68\% and 90\% C.L.~intervals are also shown in the same figure. 

The impact of the systematic uncertainties has been studied by repeating the hypothesis test without including the systematic uncertainties in the distribution of the test statistic.  
In Appendix~\ref{app:p_value_no_systematics}, figure~\ref{fig:comparison_systematics_allMajo}, the \pvalue distributions obtained without including the systematic uncertainties in the generation of the pseudo-experiments are shown, in comparison to the distributions shown in figure~\ref{fig:majo_only}. The contribution of the systematic uncertainties to the limits is about 12\%, 15\%, 14\%, and 25\% respectively for the Majoron decay modes with spectral index $n = 1,2,3,$ and 7.

The one-sided intervals on the number of events are converted to lower limits on the half-life of the decays. The latter can be related to the neutrino-Majoron coupling constant $g_J$ through the relation:
\begin{equation}
     [T_{1/2}]^{-1} = g_J^{2m} |g_A^2 \, \mathcal{M}_\alpha|^2 \, G^\alpha \;,  
     \label{eq:majorons}
\end{equation}
where $g_A=1.27$ is the free-nucleon value of the axial vector coupling constant, $m$ is the number of emitted Majorons, $\mathcal{M}_\alpha$ is the nuclear matrix element, and $G^\alpha$ is the phase space of the decay. The nuclear matrix elements for the decay mode with $n=1$ are the same as for \onbb decay. We use the full set of available calculations, which lie in the range 2.66 -- 6.34 for \Ge~\cite{PhysRevLett.105.252503,PhysRevC.87.064302,PhysRevLett.111.142501,PhysRevC.93.024308,PhysRevC.91.024613,PhysRevC.91.034304,Menendez:2017fdf,PhysRevC.95.024305,PhysRevC.98.064325,PhysRevC.97.045503,PhysRevC.101.044315}. The nuclear matrix elements for the decay modes with $n=3$ and $n=7$ have been calculated in~\cite{Kotila_2021} and the phase space factors for all the decay modes in~\cite{PhysRevC.91.064310} -- for a compilation see table~\ref{tab:comparison_majo} in appendix~\ref{app:comparison_majorons}. For $n=2$, there are no nuclear matrix elements and phase space calculations available thus, only the limit on the half-life is given. 
For all the considered decays, table~\ref{tab:majorons_results_expected_observed} presents a summary of the observed $T_{1/2}$ limits and the associated sensitivities, as well as of the deduced neutrino-Majoron coupling constant $g_J$. 

\begin{table}
    \centering
    \begin{tabular}{cccc}
    \toprule
        Decay mode & \multicolumn{2}{c}{$T_{1/2}$ (yr)}  &  $g_J$ \\
          & Sensitivity & Observed limit &  \\
        \midrule
        \onbbM ($n=1$)  & 3.5$\cdot 10^{23}$ & > 6.4$\cdot 10^{23}$ & < (1.8 -- 4.4)$\cdot10^{-5}$ \\
        \onbbM ($n=2$)  & 2.5$\cdot 10^{23}$ & > 2.9$\cdot 10^{23}$ & -- \\
        \onbbM ($n=3$)  & 1.3$\cdot 10^{23}$ & > 1.2$\cdot 10^{23}$ & < 0.017 \\
        \onbbMM ($n=3$) & 1.3$\cdot 10^{23}$ & > 1.2$\cdot 10^{23}$ & < 1.2 \\
        \onbbMM ($n=7$) & 5.8$\cdot 10^{22}$ & > 1.0$\cdot 10^{23}$ & < 1.0 \\
        \bottomrule
    \end{tabular}
    \caption{Sensitivities and observed lower limits on the half-life and upper limits on the coupling constant at 90\% C.L. for the different Majorons-involving decay modes.}
    \label{tab:majorons_results_expected_observed}
\end{table}

The obtained limits represent an improvement of factor $\sim 2$ compared to \gerda \PI results~\cite{Agostini:2015nwa}. Table~\ref{tab:comparison_majo} in appendix~\ref{app:comparison_majorons} shows a comparison with the results from other double-$\beta$ decay isotopes~\cite{Gando2012,NEMO-3:2015jgm,Barabash:2018yjq,NEMO-3:2019gwo,Kharusi:2021jez}. For a consistent comparison, the neutrino-Majoron coupling constants have been recalculated for these isotopes using the phase space factors and nuclear matrix elements of the authors quoted above. A major deviation from the published $g_J$ values occurred for the \onbbM ($n=3$) decay mode of \Xe, where the limit improved by more than a factor of 2. 
For $n=7$, our limit on the half-life of the \Ge decay is a factor of 2 better than the current best limit~\cite{Kharusi:2021jez}, despite a factor of $\sim$7 lower exposure used in this work. This result can be attributed to the particularly low background achieved in the energy region where the bulk of the \nnbb decay distribution lies and where the deformation introduced by the $n=7$ model is expected. Nevertheless, the corresponding limits on the coupling constant are comparable since the phase space of the decay in \Ge is only 0.42$\cdot 10^{-18}$\,yr$^{-1}$, compared to 12.5$\cdot 10^{-18}$\,yr$^{-1}$ in \Xe~\cite{PhysRevC.91.064310}. Similarly, despite largely different half-life limits, comparable coupling constants are reported for the various isotopes also in the case of the other decay modes. 

\subsection{Search for Lorentz and CPT violation}\label{sec:results_lorentz}
The violation of Lorentz and CPT symmetries is an interesting feature of many BSM theories that aim to unify quantum mechanics and general relativity. The Standard Model Extension (SME) is the general framework that embodies effective quantum field operators that violate Lorentz and CPT symmetries in the SM. The size of the breaking is regulated by the respective SME coefficients. Experimental constraints on these coefficients have been set from different sectors of physics, including matter, photon, neutrino, and gravity~\cite{Kosteleck__2011}. In the neutrino sector, oscillation experiments and observation of very-high-energy neutrinos have been setting the most stringent constraints on many of the SME coefficients
~\cite{Kostelecky:2003xn,MINOS:2012ozn,Super-Kamiokande:2014exs,MiniBooNE:2011pix,IceCube:2010fyu,Diaz:2013wia}. The effects of the so-called \textit{counter-shaded} operator, which is independent of the momentum of the neutrino, escape detection through measurement of neutrino oscillations and time of flight and can be uniquely accessed by studies of weak decays~\cite{Diaz:2013saa,Diaz:2013ywa}. 

If the Lorentz symmetry is violated in the neutrino sector, the differential element of the anti-neutrino momentum modifies from its standard expression $d^3q = 4\pi \omega^2 d\omega$ to $d^3q = 4\pi ( \omega^2 + 2\alv \omega) d\omega$, where \alv is the isotropic component of the SME coefficient corresponding to the above mentioned {\it counter-shaded} operator. In the \nnbb decay, this induces a change in the total decay rate that can be written as the sum of two terms: 
\begin{equation}\label{eq:lorentz}
    \Gamma^{2\nu} = \Gamma_\text{SM} + d\Gamma_\text{LV} \;,
\end{equation}
where the first term is the SM decay rate and the second term an additional perturbation due to Lorentz violation. 
In fact, \alv only affects the kinematics of the decay, {\it i.e.} the phase space factor $G^{2\nu}$, while the nuclear matrix element $\mathcal{M}_{2\nu}$ stays the same:
\begin{equation}\label{eq:lorentz_2}
    \Gamma^{2\nu} =  |g_A^2 \, \mathcal{M}_{2\nu}|^2 \, (G^{2\nu}_{SM} + dG_{LV}) \;,
\end{equation}
where the phase space factor $dG_{LV}$ contains the coefficient \alv.
The energy distribution of the perturbation has a softer spectral index, $n=4$. This is shown in figure~\ref{fig:shape_doublebeta}, in comparison to the SM \nnbb decay distribution. The quantity $dG_{LV}/\alv$ is independent of the Lorentz violating coefficient and needs to be calculated to constrain \alv experimentally.
Defining $w$ as the ratio between the integral of the Lorentz violating perturbation's distribution and the integral of the SM \nnbb decay distribution, which is the measured quantity, this can be related to the coefficient \alv through the computed ratio $\mathcal{R} = G^{2\nu} / (dG_{LV}/\alv)$:
\begin{equation}
    \alv = w \cdot \mathcal{R}  \;.
\end{equation}
We used the calculation in~\cite{Stoica2020}, performed using exact electron wave functions for building the Fermi functions and with the inclusion of finite nuclear size and screening effects:
\begin{equation}
    \mathcal{R} = 1.19\cdot 10^{-4}\, GeV \;.
\end{equation}

In the search for Lorentz violation, the parameter of interest is the parameter $w$ defined above, which is directly proportional to the Lorentz violating coefficient \alv.

\begin{figure}
    \centering
    \includegraphics[width=0.6\textwidth]{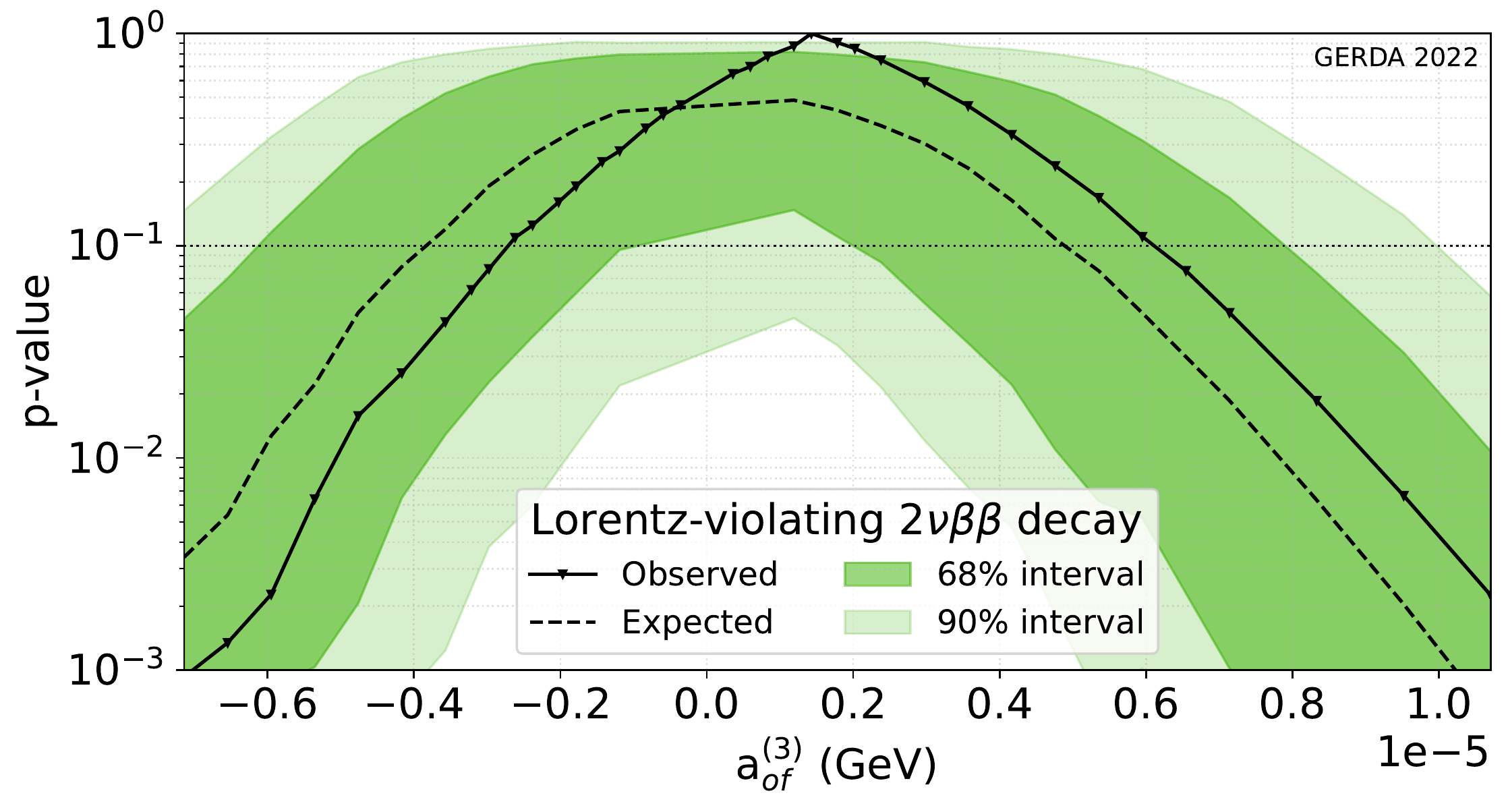}
    \caption{\pvalue as a function of the Lorentz-violating coefficient \alv.
    For details of the graphical representation, see caption of figure~\ref{fig:majo_only}.}
    \label{fig:lorentz_only}
\end{figure}

No evidence of deviation from the SM \nnbb decay distribution has been observed. A non-zero best-fit value of \alv is found, but the 68\% C.L.~interval contains zero. Therefore, a 90\% C.L.~limit has been set on \alv. Since both positive and negative values of \alv are theoretically allowed, a two-sided interval is extracted for this parameter. The observed \pvalue has been evaluated for a set of values of \alv, using the test statistic distribution computed with Monte Carlo methods. The resulting \pvalue distribution is shown in figure~\ref{fig:lorentz_only}. The median of the distribution of the expected \pvalue, which gives the experimental sensitivity, and the 68\% and 90\% C.L.~intervals are also shown in the same figure.    

In Appendix~\ref{app:p_value_no_systematics}, figure~\ref{fig:comparison_systematics_Lorentz}, the \pvalue distribution obtained without systematic uncertainties is shown, in comparison to the distribution shown in figure~\ref{fig:lorentz_only}. The impact of the systematic uncertainties on the limit is about 34\%. 

We obtained a two-sided 90\% C.L.~interval of ($-2.7 < \alv < 6.2$) $\cdot 10^{-6}$\;GeV on the SME coefficient responsible for Lorentz violation in \nnbb decay. The expected sensitivity for the \gerda experiment under no signal hypothesis is ($-3.8 < \alv < 4.9$) $\cdot 10^{-6}$\;GeV. The limit obtained in this work with \Ge \nnbb decay is competitive with the existing limits obtained with other isotopes~\cite{EXO-200:2016hbz,CUPID:2019kto,Barabash:2018yjq,NEMO-3:2019gwo}, summarized in table~\ref{tab:lorentz_comparison_2nbb_experiments}. The current best limit on \alv, of the order of $10^{-7}$\;GeV, was obtained with a very large statistic data set with the NEMO-3 experiment, with about 4.9$\cdot 10^5$ $^{100}$Mo \nnbb decay events~\cite{NEMO-3:2019gwo}. 

Stringent limits on the {\it counter-shaded} Lorentz and CPT violation can be set with $\beta$ decay experiments. In~\cite{Diaz:2013saa}, a constrain on \alv was derived using tritium $\beta$ decay published data from the Mainz and Troitsk experiments, at the level of $10^{-8}$\;GeV. Recently, the KATRIN experiment has obtained a new limit, using data collected during the first measurement campaign: $|\alv|< 3.0\cdot 10^{-8}$\,GeV at 90\% C.L.~\cite{Aker2022}. This limit is expected to further improve up to a sensitivity of $10^{-9}$\,GeV or more, with the full KATRIN exposure~\cite{Lehnert:2021tbv}. 

\begin{table}
    \centering
    \renewcommand{\arraystretch}{1.2}
    \begin{tabular}{ccc}
    \toprule
    Isotope & Limits on \alv (GeV) at 90\% C.L. & Ref. \\ \midrule
    \Ge  & ($-2.7 < \alv < 6.2$) $\cdot 10^{-6}$ & this work \\
    \Xe  & $-2.65 \cdot 10^{-5} < \alv < 7.6 \cdot 10^{-6}$ & EXO-200~\cite{EXO-200:2016hbz} \\
    $^{116}$Cd & $\alv < 4.0 \cdot 10^{-6}$ & AURORA~\cite{Barabash:2018yjq} \\
    $^{100}$Mo  & ($-4.2 < \alv < 3.5$) $\cdot 10^{-7}$ & NEMO-3~\cite{NEMO-3:2019gwo} \\
    $^{82}$Se  & $\alv < 4.1 \cdot 10^{-6}$ & CUPID-0~\cite{CUPID:2019kto} \\
    \midrule
    \multirow{2}{*}{$^3$H (single-$\beta$ decay)} & $|\alv| < 2 \cdot 10^{-8}$ & Díaz {\it et al.}~\cite{Diaz:2013saa} \\
     & $|\alv| < 3 \cdot 10^{-8}$ & KATRIN~\cite{Aker2022} \\
    \bottomrule
    \end{tabular}
    \caption{Comparison between the results obtained by different double-$\beta$ decay experiments in the search for Lorentz violation. In the last two rows, constraints obtained with tritium $\beta$ decay are reported. The first was derived in~\cite{Diaz:2013saa} using published results from the Mainz and Troitsk experiments. The second is a recent result of the KATRIN experiment~\cite{Aker2022}.}
    \label{tab:lorentz_comparison_2nbb_experiments}
\end{table}

\subsection{Search for light exotic fermions}\label{sec:results_massivefermions}

Sterile neutrinos are neutral and right-handed singlets under the SM gauge group, postulated concerning the problem of neutrino masses and dark matter~\cite{Dasgupta:2021ies}. If sterile neutrinos ($N$) exist they can be emitted in double-$\beta$ decays (\nNbb decay), $(A,Z) \rightarrow (A,Z+2) + 2e^- + \nu + N$~\cite{Bolton:2020ncv,Agostini:2020cpz}. The total decay rate can be written as the sum of two terms:
\begin{equation}\label{eq:sterile}
    \Gamma^{tot} = \cos^4\theta \cdot \Gamma^{2\nu,\text{SM}} + 2\sin^2\theta \cos^2\theta \cdot \Gamma^{N\nu} \;,
\end{equation}
where the first term is the SM \nnbb decay with two active neutrinos and the second term is the decay in which one sterile neutrino with mass $m_N$ is emitted. The parameter $\theta$ represents the mixing angle between active and sterile neutrinos.\footnote{Different works use equivalently \sinT and $|V_{eN}|^2$ to define the mixing between active and sterile neutrinos. The expression of the total decay rate in Eq.~\ref{eq:sterile} was taken from~\cite{Agostini:2020cpz}, but it is equivalent to that obtained in~\cite{Bolton:2020ncv}, where $|V_{eN}|^2$ is used and the factor 2, which appears explicitly in Eq.~\ref{eq:sterile}, is absorbed in $\Gamma^{N\nu}$.} 
The decay in which two sterile neutrinos are emitted ($\Gamma^{2N}$) would also be possible. Nevertheless, this would be strongly suppressed by a factor of $\sin^4 \theta$ and can be neglected in the experimental search~\cite{Agostini:2020cpz}. 
Like in the case of Lorentz violation, the emission of sterile neutrinos in double-$\beta$ decay only affects the kinematics of the decay, while the nuclear matrix element $\mathcal{M}_{2\nu}$ stays the same:
\begin{equation}
    \Gamma^{tot} = |g_A^2 \, \mathcal{M}_{2\nu}|^2 \, (\cos^4\theta \cdot G^{2\nu} + 2\sin^2\theta \cos^2\theta \cdot G^{N\nu}) \;,
\end{equation}
where $G^{2\nu}$ and $G^{N\nu}$ are the phase space for the emission of two active neutrinos and one active and one sterile neutrino, respectively. 

The energy distribution of the decay with the emission of a sterile neutrino is shown in figure~\ref{fig:shape_doublebeta} for a mass of 600\;keV. The endpoint of the distribution is shifted by the mass of the sterile neutrino compared to the SM \nnbb decay distribution. Given the \qbb of \Ge, a sterile neutrino with a mass up to 2\;MeV could be emitted. Nevertheless, as the mass of the emitted sterile neutrino increases, the energy distribution moves to the left, finally ending up below the energy threshold of the analysis so that no deformation of the shape is expected in the fit range. In fact, in the analysis, the emission of sterile neutrinos with masses between 100\,keV and 900\,keV was investigated.  

To search for sterile neutrinos, the energy distribution of \nnbb decay is modified according to Formula~\ref{eq:sterile}, using the calculations for $\Gamma^{N\nu}$ given in~\cite{Agostini:2020cpz}. The parameter of interest is the mixing angle \sinT. The energy distribution also depends on the mass of the sterile neutrino, $m_N$. Different masses have been analyzed independently, fixing the value of the mass in the energy distribution. 

\begin{figure}
    \centering
    \includegraphics[width=\textwidth]{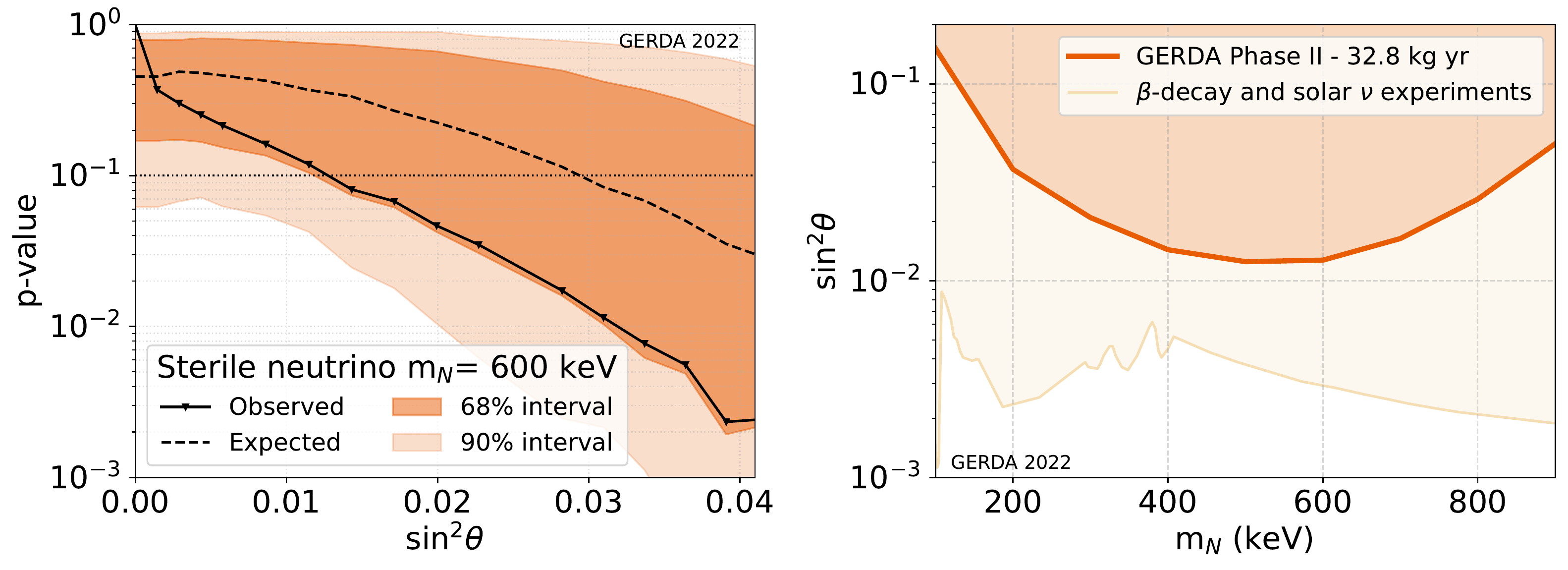}
    \caption{(left) \pvalue as a function of the mixing angle \sinT for a sterile neutrino with a mass of 600\;keV.
    For details of the graphical representation, see caption of figure~\ref{fig:majo_only}.
    (right) Results at 90\% C.L.~on the mixing angle \sinT as a function of the sterile neutrino mass. The excluded region is indicated by the orange shadowed area. Existing bounds from single-$\beta$ decay experiments~\cite{Holzschuh:2000nj,Schreckenbach:1983cg,Deutsch:1990ut,Derbin:2018dbu} and solar neutrinos~\cite{Borexino:2013bot} are also shown.}
    \label{fig:pvalues_limits_sterile}
\end{figure}

No evidence of deviation from the SM \nnbb decay distribution has been observed. The best-fit value of \sinT lies at zero for all the considered masses. Therefore, a 90\% C.L.~limit has been set on \sinT. The observed \pvalue has been evaluated for a set of values of \sinT, using the test statistic distribution computed with Monte Carlo methods. The resulting \pvalue distribution is shown in figure~\ref{fig:pvalues_limits_sterile} for sterile neutrinos with a mass of 600\;keV. The median of the distribution of the expected \pvalue, which gives the experimental sensitivity, and the 68\% and 90\% C.L.~intervals are also shown in the same figure. 

The 90\% C.L.~interval obtained on the mixing between sterile and active neutrinos is $\sinT<0.013$, for a mass of the sterile neutrino of 600\;keV. The expected sensitivity for the \gerda experiment under the no signal hypothesis is $\sinT<0.029$. Limits on the mixing \sinT and the corresponding sensitivities for the different sterile neutrino masses considered are summarized in table~\ref{tab:limits_sensitivity_sterile}. The limits on \sinT are also shown as a function of the sterile neutrino mass in figure~\ref{fig:pvalues_limits_sterile}. The best limit is obtained for a mass of 500--600\;keV. The sensitivity to smaller masses is limited by the correlation between the SM \nnbb decay and the decay with a sterile neutrino. For low masses, the distribution of the decay into sterile neutrinos becomes indistinguishable from the SM \nnbb decay distribution, where two anti-neutrino are emitted. 
Existing bounds on the sterile neutrino mixing from $\beta$ decay experiments~\cite{Holzschuh:2000nj,Schreckenbach:1983cg,Deutsch:1990ut,Derbin:2018dbu} and solar neutrinos~\cite{Borexino:2013bot} are also shown in figure~\ref{fig:pvalues_limits_sterile}. The limits obtained in this work are still not competitive, but they demonstrate the potential of double-$\beta$ decay experiments to search for sterile neutrino, with larger exposure data sets and good control of the systematic uncertainties.

The impact of the systematic uncertainties on the limits is about 17 -- 40\,\%, depending on the sterile neutrino mass. The \pvalue distribution obtained without systematic uncertainties is shown in Appendix~\ref{app:p_value_no_systematics}, figure~\ref{fig:comparison_systematics_sterile}, in comparison to the distribution shown in figure~\ref{fig:pvalues_limits_sterile} for $m_N=600$\,keV. The limits on \sinT as a function of the sterile neutrino mass obtained without systematic uncertainties are also shown in Appendix~\ref{app:p_value_no_systematics}, figure~\ref{fig:comparison_systematics_sterile}. 

\begin{table}
    \centering
    \begin{tabular}{ccc}\toprule
    m$_N$ (keV) & \multicolumn{2}{c}{\sinT} \\
     & Sensitivity & Observed limit \\ \midrule
        100 & 0.19 & < 0.15 \\
        200 & 0.056 & < 0.037 \\
        300 & 0.035 & < 0.021 \\
        400 & 0.028 & < 0.014 \\
        500 & 0.026 & < 0.013 \\
        600 & 0.029 & < 0.013 \\
        700 & 0.035 & < 0.016 \\
        800 & 0.047 & < 0.026 \\
        900 & 0.087 & < 0.050 \\ \bottomrule
    \end{tabular}
    \caption{Sensitivities and upper limits on the mixing angle \sinT between active and sterile neutrinos, for different sterile neutrino masses m$_N$. }
    \label{tab:limits_sensitivity_sterile}
\end{table}

In~\cite{Agostini:2020cpz}, an extension of the sterile neutrino model is considered, in which the neutral singlet fermion $\chi$, potentially a dark matter candidate, is introduced in the theory furnished with a discrete $\mathbb{Z}_2$ symmetry to make it stable. Such a fermion would interact with the neutrino through an effective four-fermion scalar interaction and can be produced in pair in double-$\beta$ decays (\onbbff decay), $(A,Z) \rightarrow (A,Z+2) + 2e^- + 2\chi$. Given the additional $\mathbb{Z}_2$ symmetry, such a fermion cannot be produced in single-$\beta$ decay. Hence, single-$\beta$ decay experiments that set the most stringent bounds on sterile neutrinos in the keV -- MeV mass range cannot test these kinds of models. 

The energy distribution expected for this decay is analogous to the sterile neutrino case, with the difference that the endpoint of the distribution is shifted by twice the mass of the exotic fermion since two particles are emitted. This is shown for \Ge and a mass of 300\;keV in figure~\ref{fig:shape_doublebeta}. Given the \qbb of \Ge at around 2\;MeV, only particles with mass below 1\;MeV can be produced in pairs. Nevertheless, as for the emission of sterile neutrinos, for high masses, the energy distribution moves below the energy threshold of the fit, and no shape deformation is expected in the fit range. We searched for pair production of $\chi$ fermions with masses between 100\;keV and 700\;keV. 

The energy distribution of the exotic decay is added to the fit as an independent component that constitutes the signal. The number of signal events observed in the fit range, which is proportional to the inverse of the decay half-life, is the parameter of interest in this analysis. The signal energy distribution also depends on the mass of the fermions, m$_\chi$. Different masses have been analyzed independently, fixing the value of the mass in the energy distribution. 

\begin{figure}
    \centering
    \includegraphics[width=\textwidth]{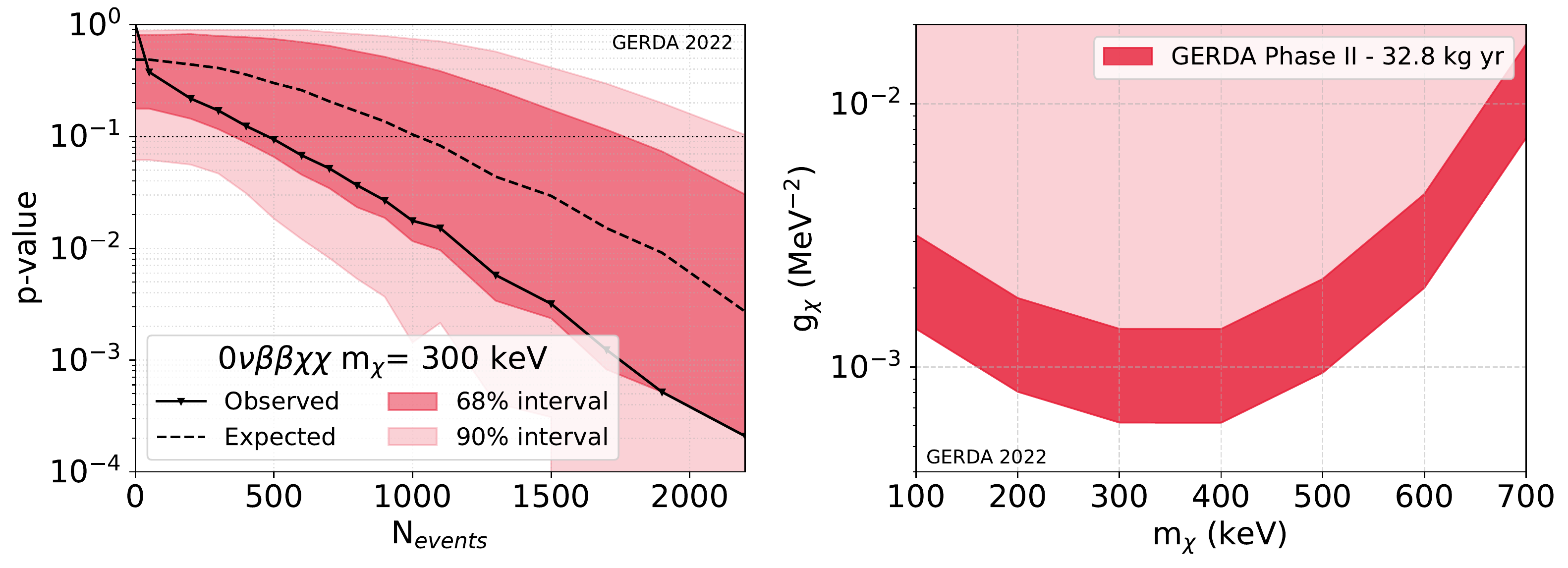}
    \caption{(left) \pvalue as a function of the number of \onbbff decay events in the fit range for a mass of the fermions of 300\;keV.
    For details of the graphical representation, see caption of figure~\ref{fig:majo_only}.
    (right) Results at 90\% C.L.~on the coupling $g_\chi$ as a function of the exotic fermion mass. The excluded region is indicated by the red shadowed area. The spread of the limits, indicated by the colored band, is given by the uncertainties on the nuclear matrix element calculations~\cite{PhysRevLett.105.252503,PhysRevC.87.064302,PhysRevLett.111.142501,PhysRevC.93.024308,PhysRevC.91.024613,PhysRevC.91.034304,Menendez:2017fdf,PhysRevC.95.024305,PhysRevC.98.064325,PhysRevC.97.045503,PhysRevC.101.044315}.}
    \label{fig:pvalue_limits_doublefermions}
\end{figure}

No evidence of a positive signal is found for any considered mass. The best fit for the number of exotic decay events is always zero for all the considered masses. Therefore, a 90\% C.L.~limit has been set on the strength of each considered decay. The observed \pvalue has been evaluated for a discrete set of values of the parameter of interest, using the test statistic distribution computed with Monte Carlo methods. The resulting \pvalue distribution is shown in figure~\ref{fig:pvalue_limits_doublefermions}, for a mass of the fermion $\chi$ of 300\;keV. The median of the distribution of the expected \pvalue, which gives the experimental sensitivity, and the 68\% and 90\% C.L~intervals are also shown in the same figure. 

The impact of the systematic uncertainties on the limits is about 10 -- 34\,\%, depending on the fermion mass. The \pvalue distribution obtained without systematic uncertainties is shown in Appendix~\ref{app:p_value_no_systematics}, figure~\ref{fig:comparison_systematics_doubleF}, for $m_\chi=300$\,keV, in comparison to the distribution shown in figure~\ref{fig:pvalue_limits_doublefermions} for the same mass. 

The one-sided intervals on the number of events are converted to lower limits on the half-life of the decays. The latter can be related to the coupling constant between the exotic fermions and neutrinos $g_\chi$ through the relation: 
 \begin{equation}
     [T_{1/2}]^{-1} = g_\chi^2 \, C \, |g_A^2\, \mathcal{M}_{0\nu}|^2 \, G^{\chi\chi} \;,  
    \label{eq:fermions}
 \end{equation}
where $C=m_e^2 / (8\pi^2R^2)$ is a constant factor that contains the electron mass $m_e$ and the nuclear radius $R=1.2\,A^{1/3}$\,fm, which in natural units (1/fm = 197.3\,MeV) is $R=0.026$\,MeV$^{-1}$ for \Ge, $\mathcal{M}_{0\nu}$ is the nuclear matrix element, that is the same as for \onbb decay, and $G^{\chi\chi}$ the phase space for the emission of two light exotic fermions. For each fermion mass, we use all the available calculations of the nuclear matrix element of \onbb decay, that are in the range 2.66 -- 6.04 for \Ge~\cite{PhysRevLett.105.252503,PhysRevC.87.064302,PhysRevLett.111.142501,PhysRevC.93.024308,PhysRevC.91.024613,PhysRevC.91.034304,Menendez:2017fdf,PhysRevC.95.024305,PhysRevC.98.064325,PhysRevC.97.045503,PhysRevC.101.044315} and compute the phase space factors according to~\cite{Agostini:2020cpz}, to convert the lower limit on the half-life in an upper limit on $g_\chi$. The lower limits on the half-life, the expected sensitivities, and the upper limits on the coupling constant at 90\% C.L. for the \onbbff decays and the different fermion masses considered are summarized in table~\ref{tab:results_massive_fermions}. In the same table, also the computed phase space factors for the different masses are reported. 

\begin{table}
    \centering
    \begin{tabular}{ccccc}\toprule
        m$_\chi$  & $G^{\chi\chi}$  & \multicolumn{2}{c}{T$_{1/2}$ (10$^{23}$\,yr)}  &  Observed $g_\chi$   \\ 
        (keV) & (10$^{-20}$\,yr$^{-1}$) & Sensitivity & Observed limit & (10$^{-3}$\,MeV$^{-2}$) \\ \midrule
        100 & 4.2  &  0.15   & > 0.18  &  < (1.4 -- 3.2)  \\
        200 & 3.3  &  0.44   & > 0.67  &  < (0.8 -- 1.8) \\
        300 & 2.3  &  0.77   & > 1.6   &  < (0.6 -- 1.4)   \\
        400 & 1.4  &  1.1    & > 2.5   &  < (0.6 -- 1.4)   \\
        500 & 0.75 &  1.2    & > 2.1   &  < (0.9 -- 2.2)  \\
        600 & 0.32 &  1.0    & > 1.1   &  < (2.0 -- 4.6)   \\
        700 & 0.10 &  0.16   & > 0.25  &  < (7.4 -- 17)  \\
        \bottomrule
    \end{tabular}
    \caption{Phase space factors $G^{\chi\chi}$, expected sensitivities and lower limits on the half-life T$_{1/2}$, and upper limits on the coupling constant $g_\chi$ at 90\% C.L., for different fermion masses m$_\chi$. }
    \label{tab:results_massive_fermions}
\end{table}

The limits on the coupling $g_\chi$ are also shown as a function of the fermions mass in figure~\ref{fig:pvalue_limits_doublefermions}. The limits obtained without systematic uncertainties are shown in Appendix~\ref{app:p_value_no_systematics}, figure~\ref{fig:comparison_systematics_doubleF}. The best limit is obtained for masses of 300 -- 400\;keV. The sensitivity to smaller masses is limited by the correlation between the energy distribution of this decay with the \nnbb decay distribution. For low masses, the distribution predicted for \onbbff decay becomes indistinguishable from the SM \nnbb decay distribution, where two anti-neutrinos are emitted.
At higher masses, the sensitivity on $g_\chi$ is weaker because of the limited space phase factor available for the decay. The width of the band is due to uncertainties in the nuclear matrix element calculations. This result represents the first experimental constraint on the pair production of two light exotic fermions in double-$\beta$ decays.

\section{Conclusions}\label{sec:conclusions}

\begin{figure}[tbp]
\centering 
\includegraphics[width=.95\textwidth]{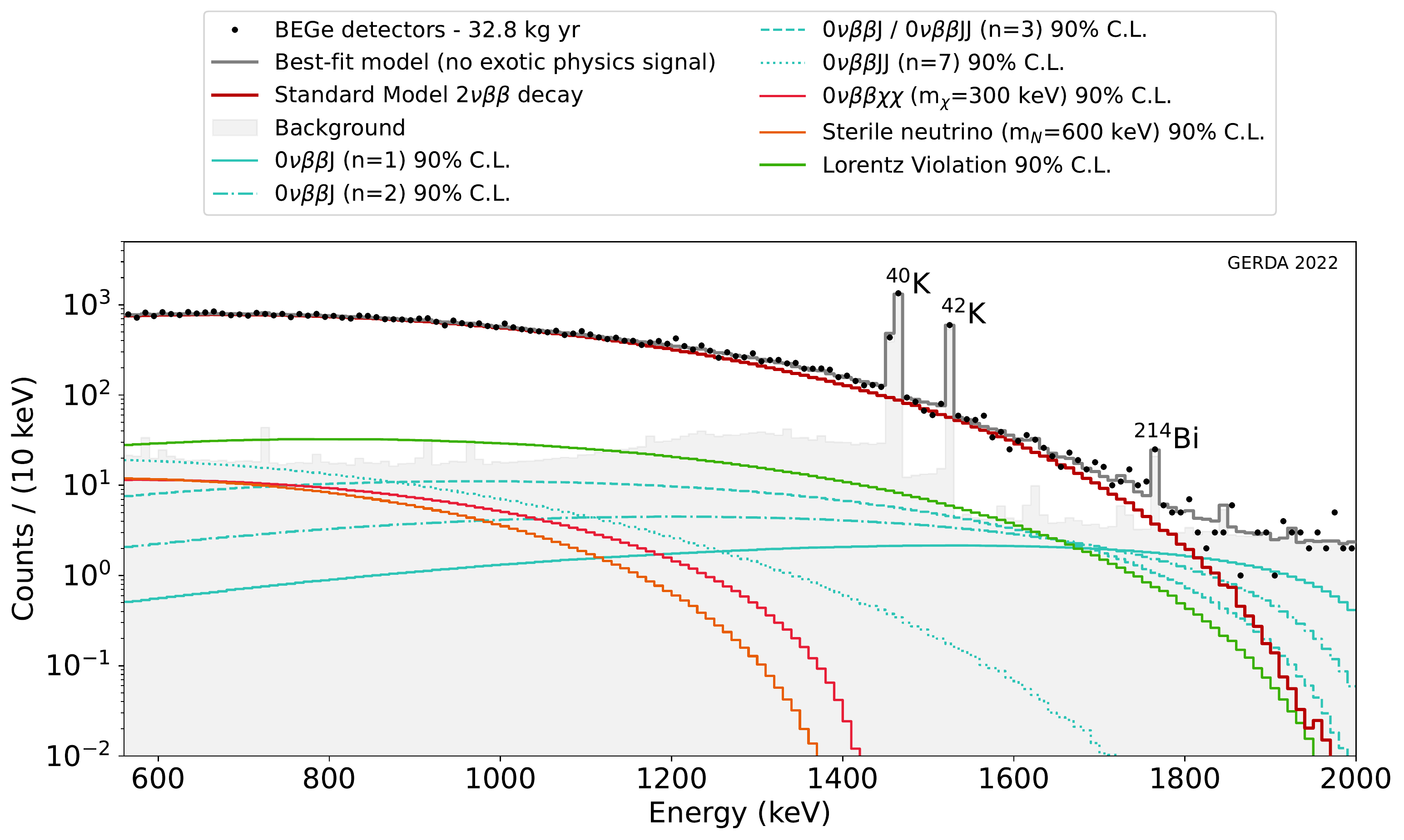}
\caption{\label{fig:all_new_physics}Data energy spectrum and best-fit model, corresponding to the absence of any new physics signal. The contributions from the SM \nnbb decay and other backgrounds are also shown separately by the solid red line and the shadowed area, respectively. The most prominent $\gamma$-lines are labeled. The 90\% C.L.~limits on the different new physics contributions obtained in this work are also visualized with different colors.}
\end{figure}

We presented the search for BSM double-$\beta$ decay modes of \Ge, performed on a selected data set collected during \PII of the \gerda experiment. No indication of deviations from the SM \nnbb decay distribution was found for any of the considered decay modes. Limits on the different models have been set with a frequentist hypothesis test based on the profile likelihood test statistic, whose distributions have been evaluated with Monte Carlo methods. 

In figure~\ref{fig:all_new_physics} the analyzed data set is shown, together with the best fit model, corresponding to the absence of any new physics signal. The contributions from the SM \nnbb decay and other backgrounds are also shown separately. The limits at 90\% C.L.~on the different new physics contributions obtained from the individual analysis are shown. All the results presented in this work represent the most stringent limits obtained with \Ge to date.

The search for \onbb decay with the emission of Majorons with \Ge was already performed in \PI of the \gerda experiment~\cite{Agostini:2015nwa}. The improvement of a factor of $\sim2$ obtained in this work, with only slightly higher exposure, can be attributed to the lower background and smaller impact of the systematic uncertainties. In this work, we searched for hints of violation of the Lorentz symmetry in \nnbb decay for the first time with \Ge and the first experimental search for the emission of light exotic fermions, including sterile neutrinos, in \nnbb decay. 

The impact of the systematic uncertainties on the limits presented here has been studied by repeating the hypothesis test, where the distributions of the test statistic do not include any systematic uncertainties. This showed that the contribution of the systematic uncertainties to the limits is about 10 -- 40\%, depending on the considered decay mode. 

The LEGEND experiment~\cite{LEGEND:2021bnm}, the future of double-$\beta$ decay physics with \Ge, will provide in the next decade a large statistic data set of \nnbb decays. A substantial improvement of the sensitivity to BSM searches still requires a further reduction of the systematic uncertainties that will become dominant as the statistics increases. For instance, a better understanding of the background model after the LAr veto cut would be beneficial. The uncertainty of the location of the background sources is indeed one of the dominant contributions to the systematic uncertainty of the results presented in this work. With increased statistics and reduction of the systematic uncertainties, the LEGEND experiment would have the possibility to test unexplored regions of the parameter space, for example, in the search for sterile neutrino with masses of hundreds of keV~\cite{Agostini:2020cpz}.

\acknowledgments
We would like to thank F.~Iachello and J.~Kotila for the useful discussions and for providing the calculations of the \nnbb energy spectrum under the different theoretical assumptions considered in this work.

The \textsc{Gerda} experiment is supported financially by
the German Federal Ministry for Education and Research (BMBF),
the German Research Foundation (DFG),
the Italian Istituto Nazionale di Fisica Nucleare (INFN),
the Max Planck Society (MPG),
the Polish National Science Centre (NCN),
the Foundation for Polish Science (TEAM/2016-2/17),
the Russian Foundation for Basic Research,
and the Swiss National Science Foundation (SNF).
This project has received funding/support from the European Union's
\textsc{Horizon 2020} research and innovation programme under
the Marie Sklodowska-Curie grant agreements No 690575 and No 674896.
This work was supported by the Science and Technology Facilities Council, part of the U.K. Research and Innovation (Grant No. ST/T004169/1).
The institutions also acknowledge internal financial support.
 
The \textsc{Gerda} collaboration thanks the directors and the staff of the LNGS for their continuous strong support of the \textsc{Gerda} experiment.

\appendix

\section{\texorpdfstring{\nnbb}{two-neutrino double-beta} decay to the \texorpdfstring{$0^+_1$}{01+} excited state}
\label{app:excited_states}
The \nnbb decay of \Ge to the $0^+_1$ excited state of $^{76}$Se is predicted with a half-life of $(2.6\pm0.13)\cdot 10^{24}$\,yr and a branching ratio (b.r.) of 0.074\%~\cite{Kostensalo2022}. This transition has not been observed in \gerda data. A lower limit on the half-life was set with \gerda \PI data: T$_{1/2} > 3.7\cdot 10^{23}$\,yr at 90\% C.L.~\cite{GERDA2015}. A more stringent limit was set by the \textsc{Majorana Demonstrator}: T$_{1/2} > 7.5\cdot 10^{23}$\,yr at 90\% C.L.~\cite{MAJORANA2020}.

\begin{figure}
    \centering
    \includegraphics[width=0.8\textwidth]{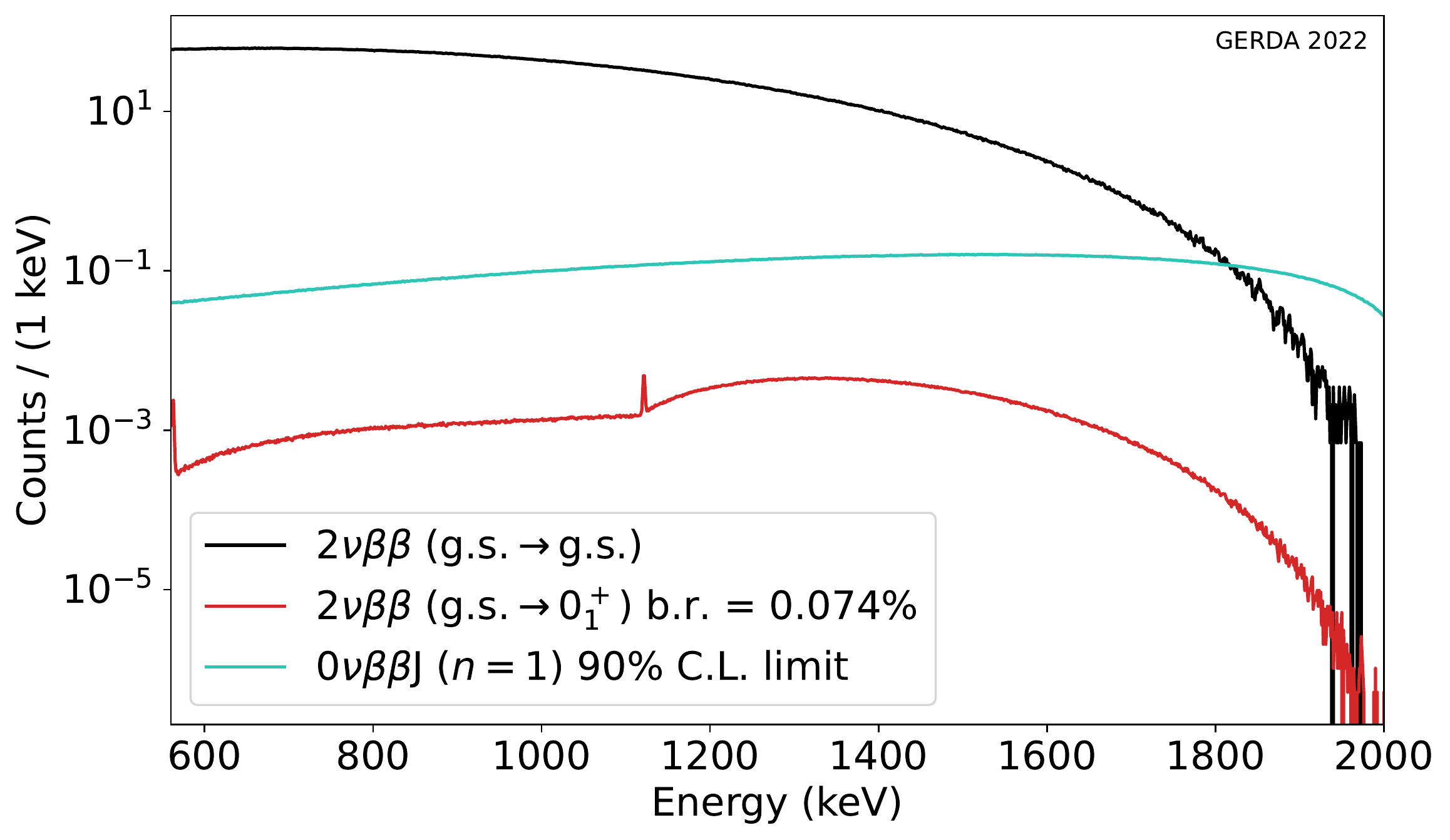}
    \caption{Simulated \nnbb decay energy distributions in the BEGe detectors after detector anti-coincidence cut and LAr veto cut. The transition to the ground state is shown in black and normalized to the number of events extracted from the fit in this work. The transition to the $0^+_1$ excited state is shown in red and normalized considering the predicted b.r. For comparison, the energy distribution of the $0\nu\beta\beta$J decay ($n=1$) normalized to the 90\% C.L. limit is also shown.}
    \label{fig:excited_states}
\end{figure}

Here, we discuss the possible systematic uncertainty that the existence of the transition to the $0^+_1$ excited state with a half-life of the order of $10^{24}$\,yr would introduce in the search for exotic double-$\beta$ decay modes. The transition to the $0^+_1$ excited state is followed by the emission of two $\gamma$s with energies of 563.2\,keV and 559.1\,keV. Given the decay in one germanium detector, the two $\gamma$s can be detected in the same detector, travel in the detector array, and be detected in one of the surrounding germanium detectors, or escape the detectors and deposit their energy in the nearby LAr volume. Only in the first case, the event topology is the same as the \nnbb decay to the ground state, and the event would pass the data selection. To estimate the detection efficiency of the $0^+_1$ transition relative to the ground state transition, we used Monte Carlo simulations performed with the \mage framework~\cite{Boswell:2010mr}. After detector anti-coincidence cut and LAr veto cut, the efficiency of the $0^+_1$ transition, relative to the ground state transition, is $\epsilon=8.88\%$ in the energy range of the analysis. Given the number of events associated with the \nnbb decay to the ground state obtained from the fit (N$_{\nnbb,\,\text{g.s.}} = 46430$\,events), and the predicted b.r., we can calculate the number of expected events in the analysis range from the \nnbb decay to the $0^+_1$ excited state:
\begin{equation}
    \text{N}_{\nnbb,\, 0^+_1} = \text{N}_{\nnbb,\,\text{g.s.}} \cdot \epsilon \cdot \text{b.r.} = 46430 \cdot 0.0888 \cdot 0.00074 \simeq 3 \, \text{events}\;.
\end{equation}
In figure~\ref{fig:excited_states}, the simulated \nnbb decay energy distribution in the BEGe detectors after anti-coincidence cut and LAr veto cut is shown for the ground state transition and the $0^+_1$ excited state transition. The first is normalized to the number of events obtained from the fit in the analysis range, the second is normalized by taking into account the predicted b.r. 

To estimate the systematic uncertainty that ignoring such a contribution in the background model would introduce in the analysis, we consider, for instance, the Majoron-involving decay mode with $n=1$. The 90\% C.L. limit obtained on the number of $0\nu\beta\beta$J decay events lies at 210 events. The number of expected events from the $0^+_1$ excited state transition is, thus, only $\sim 1\%$ of the number of $0\nu\beta\beta$J decay events from which the 90\% C.L. limit is extracted. For this reason, the exclusion of the \nnbb decay to the $0^+_1$ excited state from the background model is retained to have a negligible impact on the limits obtained in this work.

\section{Impact of systematic uncertainties}
\label{app:p_value_no_systematics}

\begin{figure}[!]
    \centering
    \includegraphics[width=\textwidth]{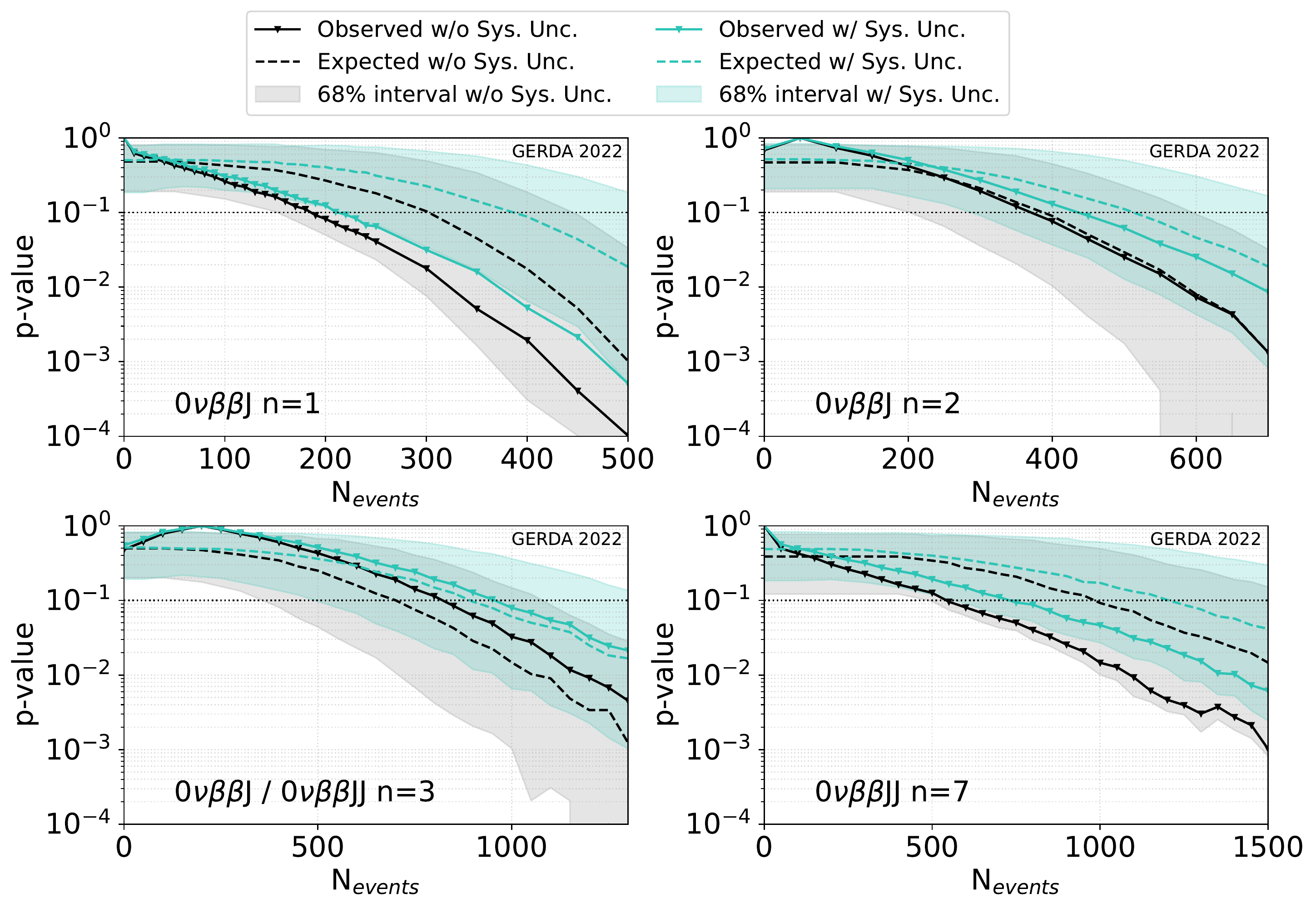}
    \caption{\pvalue as a function of the number of events in the fit range for the indicated Majorons decay modes obtained without including the systematic uncertainties in the test statistic distributions. The median of the \pvalue distribution for the \gerda experiment under the no signal hypothesis and the observed \pvalue for the \gerda data are shown by the black continuous and dashed lines, respectively. The spread of the \pvalue distribution, given by the central 68\% probability interval, is also shown by the gray band. For comparison, the same curves obtained with all the systematic uncertainties included in the test statistic distributions are also shown by the colored lines.}
    \label{fig:comparison_systematics_allMajo}
\end{figure}

\begin{figure}[!]
    \centering
    \includegraphics[width=0.8\textwidth]{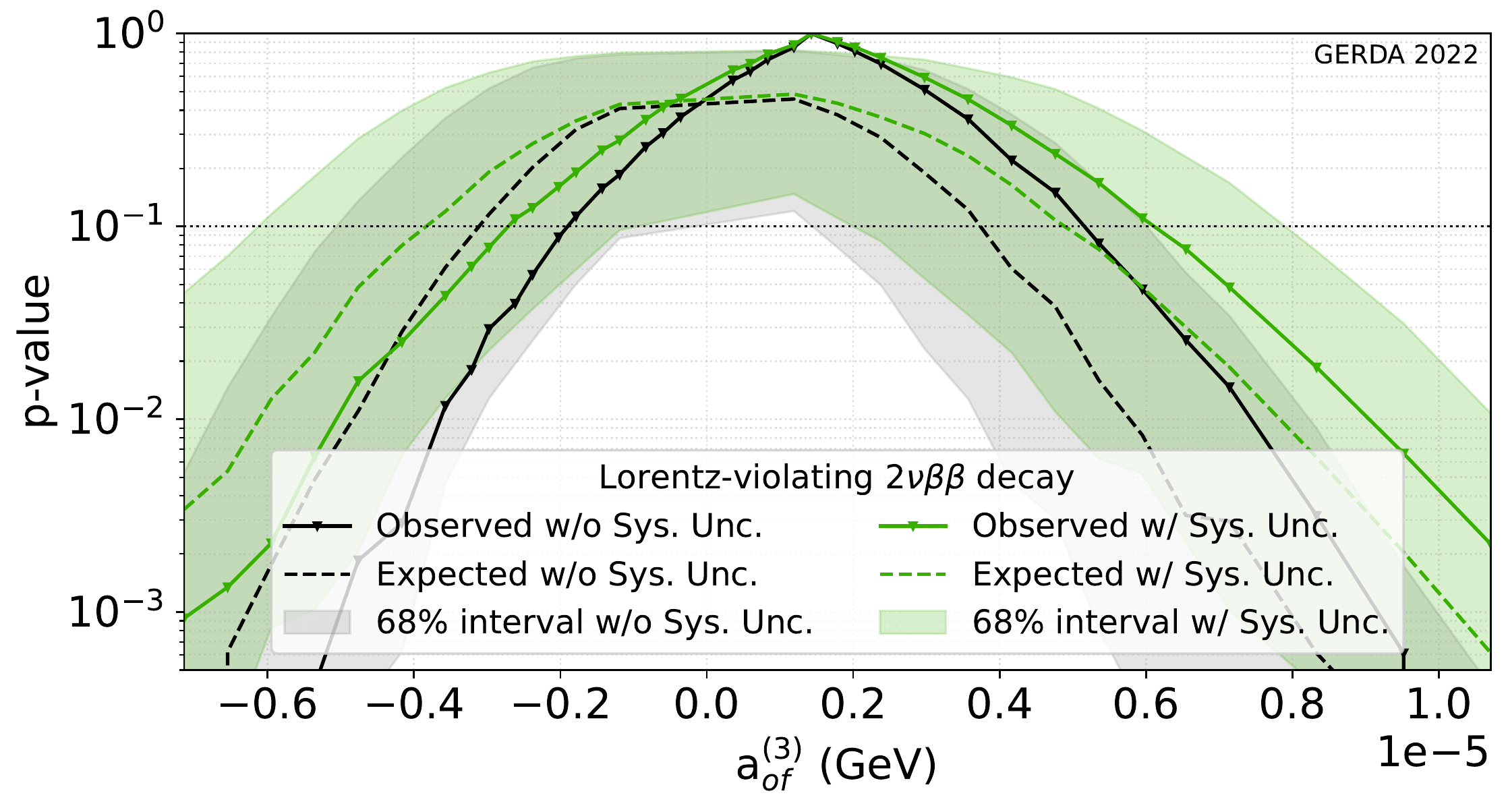}
    \caption{\pvalue as a function of the Lorentz violating coefficient \alv obtained without including the systematic uncertainties in the test statistic distributions. 
    For details of the graphical representation see caption of figure~\ref{fig:comparison_systematics_allMajo}.
    }
    \label{fig:comparison_systematics_Lorentz}
\end{figure}

\begin{figure}[!]
    \centering
    \includegraphics[width=\textwidth]{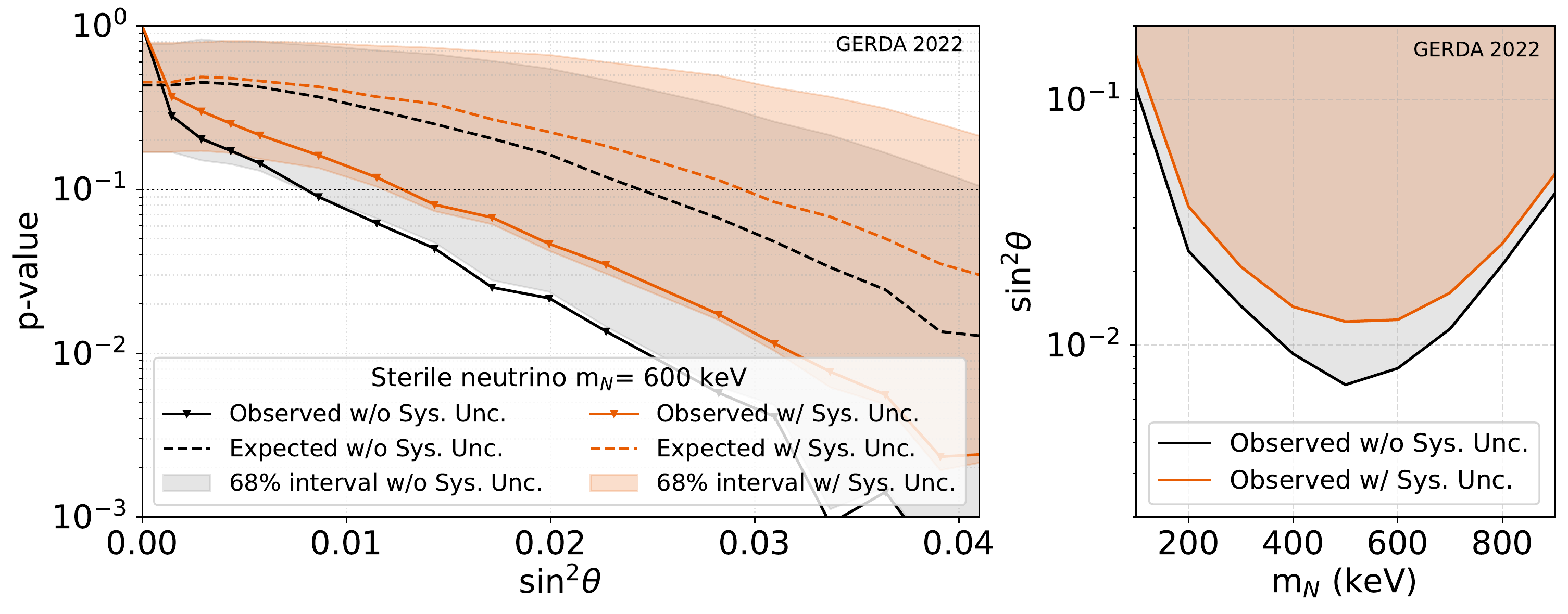}
    \caption{(left) \pvalue as a function of the mixing \sinT obtained for a sterile neutrino mass of 600\;keV without including the systematic uncertainties in the test statistic distributions. 
    For details of the graphical representation see caption of figure~\ref{fig:comparison_systematics_allMajo}.
    (right) Comparison of the results at 90\% C.L.~on the mixing angle \sinT as a function of the sterile neutrino mass obtained when the systematic uncertainties are included (colored line) or not included (black line). The excluded region is indicated by the shadowed areas. }
    \label{fig:comparison_systematics_sterile}
\end{figure}

\begin{figure}[!]
    \centering
    \includegraphics[width=\textwidth]{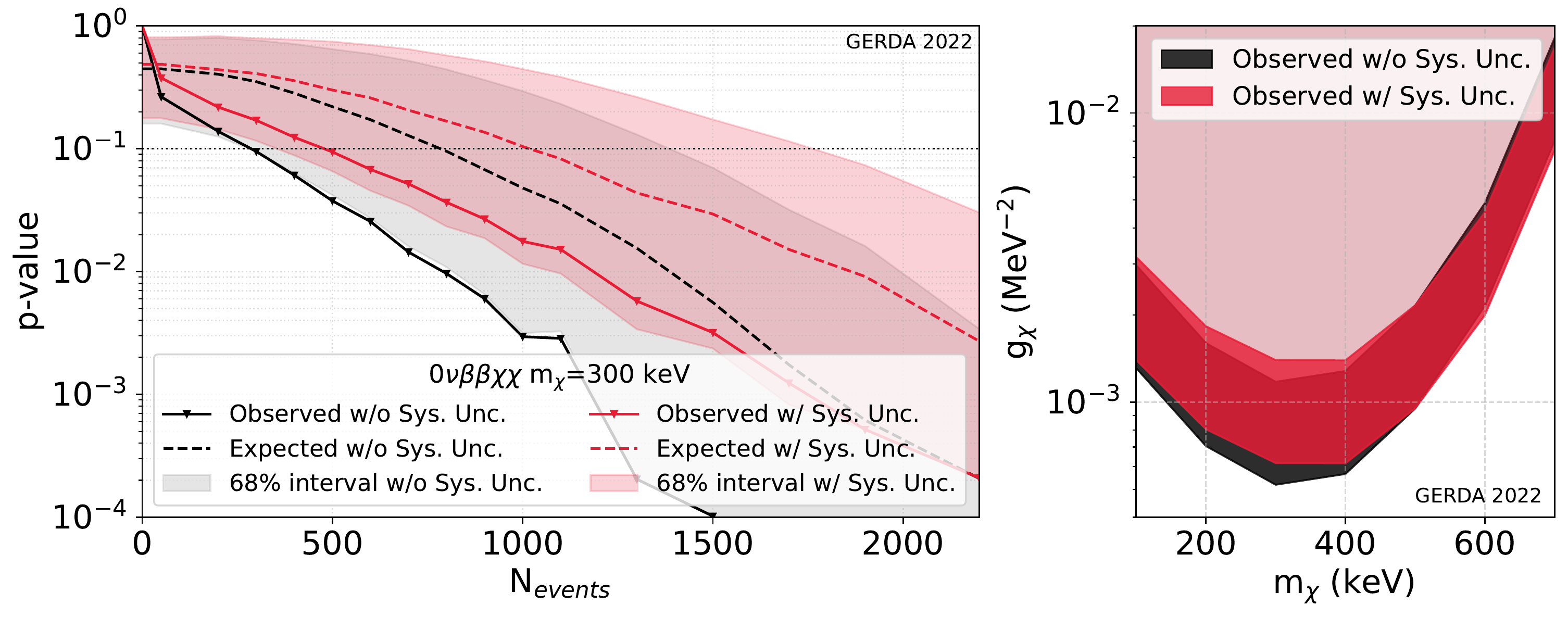}
    \caption{(left) \pvalue as a function of the number of events in the fit range for \onbbff decay obtained for a mass of the fermion $\chi$ of 300\;keV without the inclusion of the systematic uncertainties in the test statistic distributions. 
    For details of the graphical representation see caption of figure~\ref{fig:comparison_systematics_allMajo}.
    (right) Comparison of the results at 90\% C.L.~on the coupling $g_\chi$ as a function of the exotic fermion mass obtained when the systematic uncertainties are included (colored band) or not included (black band). The spread of the bands in both cases is given by the uncertainties on the nuclear matrix element calculations~\cite{PhysRevLett.105.252503,PhysRevC.87.064302,PhysRevLett.111.142501,PhysRevC.93.024308,PhysRevC.91.024613,PhysRevC.91.034304,Menendez:2017fdf,PhysRevC.95.024305,PhysRevC.98.064325,PhysRevC.97.045503,PhysRevC.101.044315}. The excluded region is indicated by the shadowed areas.}
    \label{fig:comparison_systematics_doubleF}
\end{figure}

The impact of the systematic uncertainties on the limits presented in this work has been studied. To do so, the hypothesis test has been repeated using distributions of the test statistic calculated without including the systematic uncertainties in the generation of the pseudo-experiments. A comparison between the \pvalue distributions obtained in the two cases is shown in figures~\ref{fig:comparison_systematics_allMajo},~\ref{fig:comparison_systematics_Lorentz},~\ref{fig:comparison_systematics_sterile},~\ref{fig:comparison_systematics_doubleF}, respectively for the Majoron decay modes, Lorentz violation, sterile neutrino with $m_N=600$\,keV, and \onbbff decay with $m_\chi=300$\,keV. 

\section{Comparison of Majoron results from this and other works}\label{app:comparison_majorons}

In table~\ref{tab:comparison_majo}, a comparison between the results obtained by different double-$\beta$ decay experiments, with different isotopes, in the search for double-$\beta$ decay with the emission of Majorons is presented.

\begin{table}[!]
    \centering
    \begin{tabular}{cccccc}
    \toprule
        Decay mode / & $T_{1/2}$  & Ref. & $G$ & NME & $g_J$ \\
        Isotope & (yr) & & $\,(10^{-18}\,\text{yr}^{-1})$ & & \\
        \midrule 
        \addlinespace
        \multicolumn{2}{c}{\onbbM ($n=1$)} \\ \addlinespace
        \Ge & > 6.4$\cdot 10^{23}$ & present                 & 44.2 & (2.66 -- 6.64) & < (1.8 -- 4.4)$\cdot10^{-5}$ \\
        \Xe & > 2.6$\cdot 10^{24}$ & \cite{Gando2012}        & 409 & (1.11 -- 4.77)  & < (0.4 -- 1.7)$\cdot10^{-5}$ \\
        \Xe & > 4.3$\cdot 10^{24}$ & \cite{Kharusi:2021jez}  & 409 & (1.11 -- 4.77)  & < (0.3 -- 1.3)$\cdot10^{-5}$ \\
 $^{100}$Mo & > 4.4$\cdot 10^{22}$ & \cite{NEMO-3:2015jgm}   & 598 & (3.84 -- 6.59)  & < (1.8 -- 3.1)$\cdot10^{-5}$ \\
 $^{116}$Cd & > 8.2$\cdot 10^{21}$ & \cite{Barabash:2018yjq} & 569 & (3.105 -- 5.43) & < (5.3 -- 9.2)$\cdot10^{-5}$ \\ 
        \addlinespace
        \multicolumn{2}{c}{\onbbM ($n=2$)}  \\ \addlinespace
        \Ge & > 2.9$\cdot 10^{23}$ & present                 & -- & -- & --  \\
        \Xe & > 1.0$\cdot 10^{24}$ & \cite{Gando2012}        & -- & -- & --  \\
        \Xe & > 9.8$\cdot 10^{23}$ & \cite{Kharusi:2021jez}  & -- & -- & --  \\
 $^{100}$Mo & > 9.9$\cdot 10^{21}$ & \cite{NEMO-3:2019gwo}   & -- & -- & --  \\
 $^{116}$Cd & > 4.1$\cdot 10^{21}$ & \cite{Barabash:2018yjq} & -- & -- & --  \\
        \addlinespace
        \multicolumn{2}{c}{\onbbM ($n=3$)}  \\ \addlinespace 
        \Ge & > 1.2$\cdot 10^{23}$ & present                 & 0.073 & 0.381 & < 1.7 $\cdot10^{-2}$    \\
        \Xe & > 4.5$\cdot 10^{23}$ & \cite{Gando2012}        & 1.47  & 0.160 & < 0.47 $\cdot10^{-2}$   \\
        \Xe & > 6.3$\cdot 10^{23}$ & \cite{Kharusi:2021jez}  & 1.47  & 0.160 & < 0.40 $\cdot10^{-2}$   \\
 $^{100}$Mo & > 4.4$\cdot 10^{21}$ & \cite{NEMO-3:2019gwo}   & 2.42  & 0.263 & < 2.3 $\cdot10^{-2}$    \\
 $^{116}$Cd & > 2.6$\cdot 10^{21}$ & \cite{Barabash:2018yjq} & 2.28  & 0.144 & < 5.6 $\cdot10^{-2}$    \\ 
        \addlinespace
        \multicolumn{2}{c}{\onbbMM ($n=3$)} \\ \addlinespace  
        \Ge & > 1.2$\cdot 10^{23}$ & present                 & 0.22 & 0.0026   & < 1.21   \\
        \Xe & > 4.5$\cdot 10^{23}$ & \cite{Gando2012}        & 3.05 & 0.0011   & < 0.69  \\
        \Xe & > 6.3$\cdot 10^{23}$ & \cite{Kharusi:2021jez}  & 3.05 & 0.0011   & < 0.64  \\
 $^{100}$Mo & > 4.4$\cdot 10^{21}$ & \cite{NEMO-3:2019gwo}   & 6.15 & 0.0019   & < 1.41   \\
 $^{116}$Cd & > 2.6$\cdot 10^{21}$ & \cite{Barabash:2018yjq} & 5.23 & 0.000945 & < 2.37   \\ 
        \addlinespace
        \multicolumn{2}{c}{\onbbMM ($n=7$)} \\ \addlinespace 
        \Ge & > 1.0$\cdot 10^{23}$ & present                 & 0.42 & 0.0026   & < 1.08   \\
        \Xe & > 1.1$\cdot 10^{22}$ & \cite{Gando2012}        & 12.5 & 0.0011   & < 1.23  \\
        \Xe & > 5.1$\cdot 10^{22}$ & \cite{Kharusi:2021jez}  & 12.5 & 0.0011   & < 0.84  \\
 $^{100}$Mo & > 1.2$\cdot 10^{21}$ & \cite{NEMO-3:2019gwo}   & 50.8 & 0.0019   & < 1.15  \\
 $^{116}$Cd & > 8.9$\cdot 10^{20}$ & \cite{Barabash:2018yjq} & 33.9 & 0.000945 & < 1.94   \\
        \bottomrule
    \end{tabular}
    \caption{Present results of the search for single (\onbbM) and double (\onbbMM) Majoron-involving decays of \Ge in comparison with results for the other indicated isotopes. The lower limits on the half-life $T_{1/2}$ are converted into upper limits on the neutrino-Majoron coupling constant $g_J$ using equation~\ref{eq:majorons} with the axial vector coupling constant $g_A = 1.27$ and the phase space factors $G$ from~\cite{PhysRevC.91.064310}. The nuclear matrix element (NME) calculations for the spectral index $n=1$ are taken from~\cite{Engel:2016xgb} and references therein, and for $n=3$ and $n=7$ from~\cite{Kotila_2021}.}
    \label{tab:comparison_majo}
\end{table}

\clearpage

\bibliographystyle{JHEP}
\bibliography{references}

\end{document}